\documentclass[a4paper, 12pt]{article} 
\usepackage{times}
\usepackage{geometry}
\usepackage[utf8]{inputenc}
\usepackage{ragged2e}
\usepackage[nottoc]{tocbibind}
\usepackage{hyperref}
\usepackage{multirow,multicol}
\usepackage{comment}
\usepackage{graphicx}
\usepackage{cite}
\usepackage{tikz}
\usepackage{longtable,lscape}
\usepackage{ulem}
\usepackage{pifont}
\usepackage{amssymb}
\usepackage{enumerate}
\usepackage{enumitem}
\usepackage{subfiles}
\usepackage{float}
\usepackage{longtable}
\usepackage{caption}
\usepackage{threeparttablex}
\usepackage{rotating}
\usepackage{hyperref}

\usepackage{graphicx}
\usepackage{caption}
\usepackage{subcaption}
\usepackage{xcolor}
\usepackage{soul}
\usepackage{url} 
\usepackage{pstricks-add}
\usepackage{siunitx}
\usepackage{pgfplots}
\usepackage{authblk}
\usepackage{diagbox}
\usepackage{siunitx,booktabs,caption}
\usepackage{array,makecell}
\usetikzlibrary{patterns}
\tikzset{ 
  treenode/.style = {shape=rectangle, rounded corners,
                     draw, align=center,
                     top color=white,
                     bottom color=blue!20},
  root/.style     = {treenode, font=\ttfamily\small, 
                     bottom color=red!30},
  env/.style      = {treenode, font=\ttfamily\small, minimum height=0.4cm},
  dummy/.style    = {circle,draw}
}
\definecolor{myblue}{RGB}{50,132,163}
\definecolor{myyellow}{RGB}{245,200,95}
\definecolor{mypurple}{RGB}{111,79,125}
\definecolor{mygreen}{RGB}{157,216,103}
\definecolor{myred}{RGB}{201,70,48}
\definecolor{myorange}{RGB}{242,160,86}
\definecolor{mylightblue}{RGB}{141,221,208}
\definecolor{mybrown}{RGB}{171,155,88}
\definecolor{mydarkgreen}{RGB}{113,143,80}

\definecolor{myblack}{RGB}{0,0,0}

\makeatletter

\tikzstyle{chart}=[
    legend label/.style={font={\scriptsize},anchor=west,align=left},
    legend box/.style={rectangle, draw, minimum size=5pt},
    axis/.style={black,semithick,->},
    axis label/.style={anchor=east,font={\tiny}},
]

\tikzstyle{bar chart}=[
    chart,
    bar width/.code={
        \pgfmathparse{##1/2}
        \global\let\bar@w\pgfmathresult
    },
    bar/.style={very thick, draw=white},
    bar label/.style={font={\bf\small},anchor=north},
    bar value/.style={font={\footnotesize}},
    bar width=.75,
]

\tikzstyle{pie chart}=[
    chart,
    slice/.style={line cap=round, line join=round, very thick, draw=white},
    pie title/.style={font={\bf\scriptsize}, anchor=north},
    slice type/.style n args={3}{
        ##1/.style={pattern color=##2, pattern=##3, font={\bf\scriptsize}},
        values of ##1/.style={font={\bf\scriptsize}}
    }
]

\pgfdeclarelayer{background}
\pgfdeclarelayer{foreground}
\pgfsetlayers{background,main,foreground}

\newcommand{\pie}[3][]{
    \begin{scope}[#1]
    \pgfmathsetmacro{\curA}{90}
    \pgfmathsetmacro{\r}{1}
    \def\c{(0,0)}
    \node[pie title] at (90:1.3) {#2};
    \foreach \v/\s in{#3}{
        \pgfmathsetmacro{\deltaA}{\v/100*360}
        \pgfmathsetmacro{\nextA}{\curA + \deltaA}
        \pgfmathsetmacro{\midA}{(\curA+\nextA)/2}

        \path[slice,\s] \c
            -- +(\curA:\r)
            arc (\curA:\nextA:\r)
            -- cycle;
        \pgfmathsetmacro{\d}{max((\deltaA * -(.5/50) + 1) , .5)}

        \begin{pgfonlayer}{foreground}
        \path \c -- node[pos=\d,pie values,values of \s]{$\v\%$} +(\midA:\r);
        \end{pgfonlayer}

        \global\let\curA\nextA
    }
    \end{scope}
}

\newcommand{\legend}[2][]{
    \begin{scope}[#1]
    \path
        \foreach \n/\s in {#2}
            {
                  ++(0,-10pt) node[\s,legend box] {} +(5pt,0) node[legend label] {\n}
            }
    ;
    \end{scope}
}

\newcolumntype{C}[1]{>{\centering\arraybackslash}p{#1}}

\newcounter{phase}[section]
\renewcommand{\thephase}{\Roman{phase}}
\newenvironment{phase}[2][]
{\refstepcounter{phase}
   \noindent\textbf{\texttt{Phase~\thephase .(#2)}} \rmfamily}
   
\newcounter{runexample}[section]
\newenvironment{runexample}[1][]
{\refstepcounter{runexample}\par\medskip
   \textit{Running Example:} \rmfamily}{\medskip}
   
\newcounter{observation}[section]
\newenvironment{observation}[2][]
{\refstepcounter{observation}\par\medskip
   \textbf{$^{#2}$Observation~\thesection.\theobservation:~}}

\newcounter{indication}[section]
\newenvironment{indication}[2][]
{\refstepcounter{indication}\par\medskip
  \textbf{$^{#2}$Indication~\thesection.\theindication:~}}

\title{Using Deep Learning to Solve Computer Security Challenges: A Survey\footnote{The authors of this paper are listed in alphabetic order.}}

\author[2,1]{Yoon-Ho Choi}
\author[1]{Peng Liu\thanks{Corresponding author:~pxl20@psu.edu}}
\author[1]{Zitong Shang}
\author[1]{Haizhou Wang}
\author[1]{Zhilong Wang}
\author[1]{Lan Zhang}
\author[3,1]{Junwei Zhou}
\author[1]{Qingtian Zou}

\affil[1]{The Pennsylvania State University, United States}
\affil[2]{Pusan National University, Republic of Korea}
\affil[3]{Wuhan University of Technology, China}
\date{\vspace{-5ex}}

\begin{document}
\maketitle

\begin{abstract}
Although using machine learning techniques to solve computer security challenges is not a new idea, the rapidly emerging Deep Learning technology has recently triggered a substantial amount of interests in the computer security community. This paper seeks to provide a dedicated review of the very recent research works on using Deep Learning techniques to solve computer security challenges. In particular, the review covers eight computer security problems being solved by applications of Deep Learning: security-oriented program analysis, defending return-oriented programming (ROP) attacks, achieving control-flow integrity (CFI), defending network attacks, malware classification, system-event-based anomaly detection, memory forensics, and fuzzing for software security.  
\end{abstract}

\section{Introduction}
Using machine learning techniques to solve computer security challenges is not a new idea. For example, in the year of 1998, Ghosh and others in \cite{GWC98} proposed to train a (traditional) neural network based anomaly detection scheme(i.e., detecting anomalous and unknown intrusions against programs); in the year of 2003, Hu and others in \cite{HLV03} and Heller and others in \cite{HSK03} applied Support Vector Machines to based anomaly detection scheme (e.g., detecting anomalous Windows registry accesses).  

The machine-learning-based computer security research investigations during 1990-2010, however, have not been very impactful. For example, to the best of our knowledge, none of the machine learning applications proposed in \cite{GWC98,HLV03,HSK03} has been incorporated into a  widely deployed intrusion-detection commercial product. 

Regarding why not very impactful, although researchers in the computer security community seem to have different opinions, the following remarks by Sommer and Paxson \cite{SP10} (in the context of intrusion detection) have resonated with many researchers: 
\begin{itemize}
    \item Remark A: ``It is crucial to have a clear picture of what problem a system targets: what specifically are the attacks to be detected? The more narrowly one can define the target activity, the better one can tailor a detector to its specifics and reduce the potential for misclassifications." \cite{SP10}   
    
    \item Remark B: ``If one cannot make a solid argument for the relation of the features to the attacks of interest, the resulting study risks foundering on serious flaws." \cite{SP10}  
\end{itemize}

{\color{myblack} These insightful remarks, though well aligned with the machine learning techniques used by security researchers during 1990-2010, could become a less significant concern with Deep Learning (DL), 
a rapidly emerging machine learning technology, due to the following observations. 
First, Remark A implies that even if the same machine learning method is used, 
one algorithm employing a cost function that is based on a more specifically defined target attack activity could perform substantially better than another algorithm deploying a less specifically defined cost function. 
This could be a less significant concern with DL, since a few recent studies have shown that even if the target attack activity is not narrowly defined, 
a DL model could still achieve very high classification accuracy. 
Second, Remark B implies that if feature engineering is not done properly, the trained machine learning models could be plagued by serious flaws. 
This could be a less significant concern with DL, since many deep learning neural networks require less feature engineering than conventional machine learning techniques.
}

As stated in \cite{NSCAI19}, ``DL is a statistical technique that exploits large quantities of data as training sets for a network with multiple hidden layers, called a deep neural network (DNN). A DNN is trained on a dataset, generating outputs, calculating errors, and
adjusting its internal parameters. Then the process is repeated hundreds of thousands of times until the network achieves an acceptable level of performance. It has proven to be an effective technique for image classification, object detection, speech recognition, and
natural language processing––problems that challenged researchers for decades. By learning from data, DNNs can solve some problems much more effectively, and also solve problems that were never solvable before."

Now let's take a high-level look at how DL could 
make it substantially easier to overcome the challenges 
identified by Sommer and Paxson \cite{SP10}. 
First, one major advantage of DL is that it makes 
learning  algorithms  less  dependent on  feature  engineering. This characteristic of DL makes it easier to 
overcome the challenge indicated by Remark B. 
Second, another major advantage of DL is that it 
could achieve high classification accuracy with 
minimum domain knowledge. This characteristic of DL makes it easier to overcome the challenge indicated by Remark A. 

\vspace*{2mm} 
\textbf{Key observation. } \textit{The above discussion indicates that DL could be a game changer in applying machine learning techniques to solving computer security challenges. }

\vspace*{2mm} 
Motivated by this observation, this paper seeks to provide a \textit{dedicated} review of the very recent research works on using Deep Learning techniques to solve computer security challenges. It should be noticed that since this paper aims to provide a dedicated review, non-deep-learning techniques and their security applications are out of the scope of this paper.  

The remaining of the paper is organized as follows. In Section~\ref{sec:workflow}, we present a four-phase workflow framework which we use to summarize the existing works in a unified manner. In Section~\ref{sec:programanalysis}-\ref{sec:fuzzing}, we provide a review of eight computer security problems being solved by applications of Deep Learning, respectively. In Section~\ref{sec:dis}, we will discuss certain similarity and certain dissimilarity among the existing works. In Section~\ref{sec:fur}, we mention four further areas of investigation. In Section~\ref{sec:con}, we conclude the paper. 

\section{A four-phase workflow framework can summarize the existing works in a unified manner}
\label{sec:workflow}
We found that a four-phase workflow framework can provide a unified way to summarize all the research works surveyed by us. In particular, we found that each work surveyed by us employs a particular workflow when using machine learning techniques to solve a computer security challenge, and we found that each workflow consists of two or more phases. 
By ``a unified way", we mean that every workflow surveyed by us is essentially an instantiation of a common workflow pattern which is shown in Figure \ref{fig:4phases}.    

\subsection{Definitions of the four phases}
\begin{figure}[!ht]
\begin{center}
\includegraphics[scale=0.80]{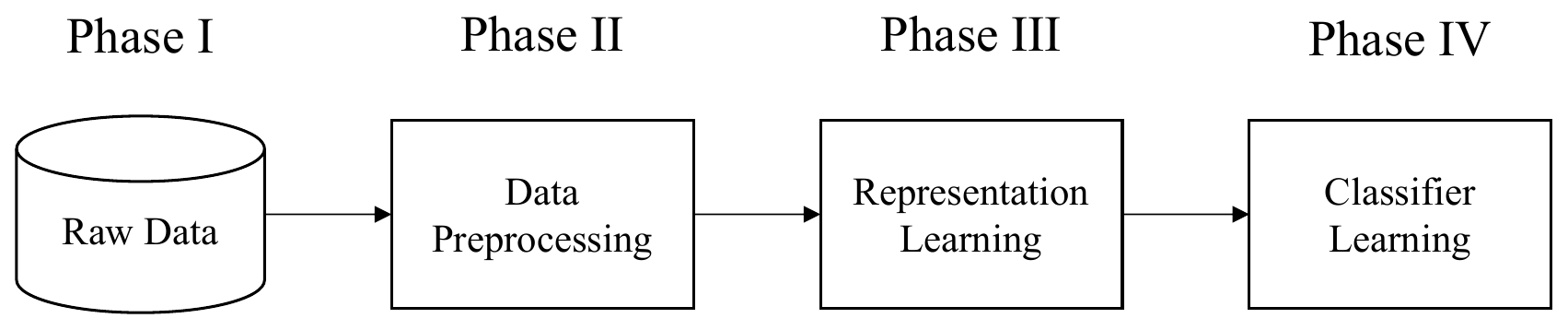}
\caption{Overview of the four-phase workflow} 
\label{fig:4phases}
\end{center}
\end{figure}

The four phases, shown in Figure \ref{fig:4phases}, are defined as follows. To make the definitions of the four phases more tangible, we use a running example to illustrate each of the four phases.

\begin{phase}{Obtaining the Raw Data}\label{phase1}

In this phase, certain raw data are collected.
\begin{runexample}
When Deep Learning is used to detect suspicious events in a Hadoop distributed file system (HDFS), the raw data are usually the events (e.g., a block is allocated, read, written, replicated, or deleted) that have happened to each block.  
Since these events are recorded in Hadoop logs, the log files hold the raw data. 
Since each event is uniquely identified by a particular (block ID, timestamp) tuple, we could simply view
the raw data as $n$ event sequences. Here $n$ is the total number of blocks in the HDFS. 
For example, the raw data collected in \cite{Xu:2009}
in total consists of 11,197,954 events. 
Since 575,139 blocks were in the HDFS, there were 
575,139 event sequences in the raw data, and on average
each event sequence had 19 events. One such event sequence is shown as follows: 

\footnotesize
\begin{verbatim}
081110 112428 31 INFO dfs.FSNamesystem: BLOCK* NameSystem.allocateBlock:
  /user/root/rand/_temporary/_task_200811101024_0001_m_001649_0/
  part-01649.blk_-1033546237298158256
081110 112428 9602 INFO dfs.DataNode$DataXceiver: 
  Receiving block blk_-1033546237298158256 src: /10.250.13.240:54015
  dest:/10.250.13.240:50010
081110 112428 9982 INFO dfs.DataNode$DataXceiver: 
  Receiving block blk_-1033546237298158256 src: /10.250.13.240:52837 
  dest:/10.250.13.240:50010
081110 112432 9982 INFO dfs.DataNode$DataXceiver: 
  writeBlock blk_-1033546237298158256 received exception
  java.io.IOException:Could not read from stream
\end{verbatim}
\normalsize
\end{runexample}
\end{phase}
\begin{phase}{Data Preprocessing}\label{phase2}

Both Phase~\ref{phase2} and Phase~\ref{phase3} aim to properly 
extract and represent the useful   
information held in the raw data collected in Phase I. 
Both Phase~\ref{phase2} and Phase~\ref{phase3} are closely related
to feature engineering. 
A key difference between Phase~\ref{phase2} and Phase~\ref{phase3} 
is that Phase~\ref{phase3} is completely dedicated to 
representation learning, while Phase~\ref{phase2} is 
focused on all the information extraction and data 
processing operations that are \textbf{not} based on 
representation learning. 
\begin{runexample}
Let's revisit the 
aforementioned HDFS. 
Each recorded event is described by unstructured text. 
In Phase~\ref{phase2}, the unstructured text
is parsed to a data structure that shows the event type and a list of event variables in (name, value) pairs. 
Since there are 29 types of events in the HDFS, 
each event is represented by an integer from 1 to 29 according to its type. 
In this way, the aforementioned example event sequence 
can be transformed to:  
  \begin{verbatim} 
   22, 5, 5, 7
  \end{verbatim}
\end{runexample}
\end{phase}

\begin{phase}{Representation Learning}\label{phase3}

As stated in \cite{Bengio2013}, ``Learning representations of the data that make it easier to extract useful information when building classifiers or other predictors."  
\begin{runexample}
Let's revisit the same HDFS. 
Although DeepLog \cite{Du:2017:deeLog} directly employed one-hot vectors to represent the event types without representation learning,  
if we view an event type as a word in a structured language, 
one may actually use the word embedding technique to represent each event type. It should be noticed that 
the word embedding technique is a representation 
learning technique. 
\end{runexample}
\end{phase}

\begin{phase}{Classifier Learning}\label{phase4}

This phase aims to build specific classifiers or other predictors through Deep Learning. 
\begin{runexample}
Let's revisit the same HDFS.
DeepLog \cite{Du:2017:deeLog} used Deep Learning to build a stacked LSTM neural network for anomaly detection. 
For example, let's consider event sequence  \{22,5,5,5,11,9,11,9,11,9,26,26,26\} in which each 
integer represents the event type of the corresponding event in the event sequence.  
Given a window size $h$ = 4, the input sample and the output label pairs to train DeepLog will be: \{22,5,5,5 $\rightarrow$ 11 \}, \{5,5,5,11 $\rightarrow$ 9 \}, \{5,5,11,9 $\rightarrow$ 11 \}, and so forth. 
In the detection stage, DeepLog examines each individual event.  It determines if an event is treated as normal or abnormal according to whether the event's type is predicted by the LSTM neural network, given the history of event types.
If the event's type is among the top $g$ predicted types, 
the event is treated as normal; otherwise, it is treated as abnormal.
\end{runexample}
\end{phase}

\subsection{Using the four-phase workflow framework to summarize some representative research works} 

In this subsection, we use the four-phase workflow framework to summarize two representative works for each security problem. 
{\color{myblack}
System security includes many sub research topics. 
However, not every research topics are suitable to adopt deep learning-based methods due to their intrinsic characteristics. 
For these security research subjects that can combine with deep-learning, 
some of them has undergone intensive research in recent years, 
others just emerging. 
We notice that there are 5 mainstream research directions in system security. 
This paper mainly focuses on system security, so the other mainstream research directions (e.g., deepfake) are out-of-scope.
Therefore, we choose these 5 widely noticed research directions,
and 3 emerging research direction in our survey:
\begin{enumerate}
\item In security-oriented program analysis, malware classification (MC), system-event-based anomaly detection (SEAD), memory forensics (MF), and defending network attacks, deep learning based methods have already undergone intensive research. 
\item In defending return-oriented programming (ROP) attacks, Control-flow integrity (CFI), and fuzzing, deep learning based methods are emerging research topics.
\end{enumerate}
We select two representative works for each research topic in our survey.
Our criteria to select papers mainly include: 1) Pioneer (one of the first papers in this field); 
2) Top (published on top conference or journal);
3) Novelty; 4) Citation (The citation of this paper is high); 
5) Effectiveness (the result of this paper is pretty good);
6) Representative (the paper is a representative work for a branch of the research direction). 
Table~\ref{tab:reason} lists the reasons why we choose each paper, which is ordered according to their importance.
}
\begin{table*}[ht]
    \footnotesize
    \caption{List of criteria we used to choose representative work for each research topic.}
    \label{tab:reason}
  \makeatletter
  \newcommand{\widenhline}{%
      \noalign {\ifnum 0=`}\fi \hrule height 0.5pt
      \futurelet \reserved@a \@xhline
  }
  \newcolumntype{I}{!{\vrule width 1pt}}
  \makeatother
  \centering
  \small
  \begin{tabular}{cIc|c|c|c}  
  \hline  
  \hline  
  \diagbox[dir=NW]{Paper}{Order} & 1 & 2 & 3  & 4 \\ \widenhline
  \hline  
  RFBNN~\cite{shin2015recognizing} & Pioneer & Top & Novelty & Citations \\ 
  \hline  
  EKLAVYA~\cite{chua2017neural} & Top & Novelty & Citation & N/A \\  
  \hline  
  ROPNN~\cite{li2018ropnn} & Pioneer & Novelty & Effectiveness & N/A \\ 
  \hline  
  HeNet~\cite{chen2018henet} & Effectiveness & Novelty & Citation & N/A \\ 
  \hline  
  Barnum~\cite{yagemann2019barnum} & Pioneer & Novelty & N/A & N/A \\  
  \hline  
  CFG-CNN~\cite{phan2017convolutional} & Representative & N/A & N/A & N/A \\ 
  \hline  
  50b(yte)-CNN\cite{millar2018deep} & Novelty & Effectiveness & N/A & N/A \\  
  \hline  
  PCNN~\cite{zhang2019pccn}  & Novelty & Effectiveness & N/A & N/A \\  
  \hline  
  Resenberg~\cite{Rosenberg2018} & Novelty & Effectiveness & Top & Representative \\  
  \hline  
  DeLaRosa~\cite{DeLaRosa2018} & Novelty & Representative & N/A & N/A  \\ 
  \hline  
  DeepLog~\cite{Du:2017:deeLog} & Pioneer & Top & Citations & N/A \\  
  \hline  
  DeepMem~\cite{Song:2018:DeepMem} & Pioneer & Top & N/A & N/A \\   
  \hline  
  NeuZZ~\cite{shi2019neuzz} & Novelty & Top & Effectiveness & N/A \\  
  \hline  
  Learn \& Fuzz~\cite{godefroid2017learn} & Pioneer & Novelty & Top & N/A \\  
  \hline 
  \hline   
  \end{tabular}  
  \end{table*}

The summary for each paper we selected is shown in Table~\ref{Table:Summary}. 
There are three columns in the table. 
In the first column, we listed eight security problems, including security-oriented program analysis, defending return-oriented programming (ROP) attacks, control-flow integrity (CFI), defending network attacks (NA), malware classification (MC), system-event-based anomaly detection (SEAD), memory forensics (MF), and fuzzing for software security. 
In the second column, we list the very recent two representative works for each security problem. From the $3$-th to $6$-th columns, we sequentially describe how the four phases are deployed at each work. 
In the  ``Summary" column, we sequentially describe how the four phases are deployed at each work, then, we list the evaluation results for each work in terms of accuracy (ACC), precision (PRC), recall (REC), F1 score (F1), false-positive rate (FPR), and false-negative rate (FNR), respectively.

\begin{center} 
\scriptsize
\begin{ThreePartTable}
\begin{TableNotes}
    \item [1] Deep Learning metrics are often not available in fuzzing papers. Typical fuzzing metrics used for evaluations are: code coverage, pass rate and bugs.
\end{TableNotes}
\newcolumntype{C}[1]{>{\centering\arraybackslash}p{#1}}

\begin{longtable}{p{1.4cm}<{\centering}p{1.4cm}<{\centering}cccccc}
\captionsetup{width=.95\textwidth}
\caption{\footnotesize Solutions using Deep Learning for eight security problems. The metrics in the Evaluation column include accuracy (ACC), precision (PRC), recall (REC), $F_{1}$ score ($F_{1}$), false positive rate (FPR), and false negative rate (FNR).}
\label{Table:Summary}\\

\toprule \toprule
\textbf{Security Problem} & \textbf{Works} & \multicolumn{6}{c}{\textbf{Summary}} \\
\cmidrule[1pt]{1-8}
\endfirsthead

 
\multicolumn{7}{c}{{Continued on Next Page\ldots}} \\
\endfoot

\insertTableNotes 
\endlastfoot
    
    {\multirow{5}{1.5cm}{Security Oriented Program Analysis~\cite{shin2015recognizing,chua2017neural,guo2019deepvsa,xu2017neural}}} 
            & {\multirow{2}{*}{RFBNN~\cite{shin2015recognizing}}} & \multicolumn{3}{C{5cm}}{Phase I} & \multicolumn{3}{C{5cm}}{Phase II}  \\
    \cmidrule[\lightrulewidth](lr){3-8}\addlinespace[0ex]
    && \multicolumn{3}{C{5.7cm}}{Dataset comes from previous paper~\cite{184521}, consisting of 2200 separate binaries. 
                        2064 of the binaries were for Linux, obtained from the coreutils, binutils, and findutils packages. 
                        The remaining 136 for Windows consist of binaries from popular open-source projects.
                        Half of the binaries were for x86, and the other half for x86-64. } 
     & \multicolumn{3}{C{5.7cm}}{They extract fixed-length subsequences (1000-byte chunks) from code section of binaries,
                        Then, use “one-hot encoding”, which converts a
                        byte into a $\mathbb{Z}^{256}$ vector.} \\
    \cmidrule[\lightrulewidth](lr){3-8}\addlinespace[0ex]
    & & \multicolumn{2}{C{4.4cm}}{Phase III} & \multicolumn{2}{C{3.8cm}}{Phase IV} & \multicolumn{2}{C{2.8cm}}{Evaluation}  \\
     \cmidrule[\lightrulewidth](lr){3-8}\addlinespace[0ex]
    & & \multicolumn{2}{C{4.4cm}}{ N/A } 
      & \multicolumn{2}{C{3.8cm}}{ Bi-directional RNN}
      &  
        \begin{tabular}[t]{p{0.2cm}<{\centering}p{0.1cm}<{\centering}p{0cm}p{0.2cm}<{\centering}p{0.1cm}<{\centering}}
        ACC:&98.4\%&&PRE:&N/A\\
        REC:&0.97&&$F_{1}$:&0.98\\
        FPR:&N/A&&FNR:&N/A\\
        \end{tabular} 
       \\

    \cmidrule[1pt]{2-8}
    & {\multirow{2}{*}{EKLAVYA\cite{chua2017neural}}} & \multicolumn{3}{C{5cm}}{Phase I} & \multicolumn{3}{C{5cm}}{Phase II}  \\
    \cmidrule[\lightrulewidth](lr){3-8}\addlinespace[0ex]
    & & \multicolumn{3}{C{5.7cm}}{They adopted source code from previous work~\cite{shin2015recognizing} as their rawdata,
                                            then obtained two datasets by using two commonly used compilers: gcc and clang, with different optimization levels ranging from O0 to O3 for both x86 and x64. 
                                            They obtained the ground truth for the function arguments by parsing the DWARF debug information.
                                            Next, they extract functions from the binaries and remove functions which are duplicates of other functions in the dataset. 
                                            Finally, they match caller snipper and callee body.  } 
     & \multicolumn{3}{C{5.7cm}}{Tokenizing the hexadecimal value of each instruction.} \\
    \cmidrule[\lightrulewidth](lr){3-8}\addlinespace[0ex]
    & & \multicolumn{2}{C{4.4cm}}{Phase III} & \multicolumn{2}{C{3.8cm}}{Phase IV} & \multicolumn{2}{C{2.8cm}}{Evaluation}  \\
    \cmidrule[\lightrulewidth](lr){3-8}\addlinespace[0ex]
    & & \multicolumn{2}{C{4.4cm}}{ Word2vec technique to compute word embeddings. } 
      & \multicolumn{2}{C{3.8cm}}{RNN}
      & 
      \begin{tabular}[t]{p{0.2cm}<{\centering}p{0.1cm}<{\centering}p{0cm}p{0.2cm}<{\centering}p{0.1cm}<{\centering}}
      ACC:&81.0\%&&PRE:&N/A\\
      REC:&N/A&&$F_{1}$:&N/A\\
      FPR:&N/A&&FNR:&N/A\\
      \end{tabular} \\

    \cmidrule[0.8pt]{1-8}
    {\multirow{5}{1.5cm}{Defending Return Oriented Programming Attacks
    \cite{li2018ropnn,chen2018henet,zhang2019deepcheck}}} 
            & {\multirow{2}{*}{ROPNN \cite{li2018ropnn}  }} & \multicolumn{3}{C{5cm}}{Phase I} & \multicolumn{3}{C{5cm}}{Phase II}  \\
    \cmidrule[\lightrulewidth](lr){3-8}\addlinespace[0ex]
    && \multicolumn{3}{C{5.7cm}}{The data is a set of gadget chains obtained from existing programs. A gadget searching tool, ROPGadget is used to find available gadgets. Gadgets are chained based on whether the produced gadget chain is executable on a CPU emulator. The raw data is represented in hexadecimal form of instruction sequences. } 
        & \multicolumn{3}{C{5.7cm}}{Form one-hot vector for bytes.} \\
    \cmidrule[\lightrulewidth](lr){3-8}\addlinespace[0ex]
    
    & & \multicolumn{2}{C{4.4cm}}{Phase III} & \multicolumn{2}{C{3.8cm}}{Phase IV} & \multicolumn{2}{C{2.8cm}}{Evaluation}  \\
    \cmidrule[\lightrulewidth](lr){3-8}\addlinespace[0ex]
    & & \multicolumn{2}{C{4.4cm}}{ N/A } 
        & \multicolumn{2}{C{3.8cm}}{ 1-D CNN}
        &  
        \begin{tabular}[t]{p{0.2cm}<{\centering}p{0.1cm}<{\centering}p{0cm}p{0.2cm}<{\centering}p{0.1cm}<{\centering}}
        ACC:&99.9\%&&PRE:&0.99\\
        REC:&N/A&&$F_{1}$:&0.01\\
        FPR:&N/A&&FNR:&N/A\\
        \end{tabular} \\
    \cmidrule[1pt]{2-8}
    & {\multirow{2}{*}{HeNet \cite{chen2018henet}}} & \multicolumn{3}{C{5cm}}{Phase I} & \multicolumn{3}{C{5cm}}{Phase II}  \\
    \cmidrule[\lightrulewidth](lr){3-8}\addlinespace[0ex]
    & & \multicolumn{3}{C{5.7cm}}{Data is acquired from Intel PT, which is a processor trace tool that can log control flow data. Taken Not-Taken (TNT) packet and Target IP (TIP) packet are the two packets of interested. Logged as binary numbers, information of executed branches can be obtained from TNT, and binary executed can be obtained from TIP. Then the binary sequences are transferred into sequences of values between 0-255, called pixels, byte by byte.
    } 
        & \multicolumn{3}{C{5.7cm}}{Given the pixel sequences, slice the whole sequence and reshape to form sequences of images for neural network training.
        } \\
    \cmidrule[\lightrulewidth](lr){3-8}\addlinespace[0ex]
    & & \multicolumn{2}{C{4.4cm}}{Phase III} & \multicolumn{2}{C{3.8cm}}{Phase IV} & \multicolumn{2}{C{2.8cm}}{Evaluation}  \\
    \cmidrule[\lightrulewidth](lr){3-8}\addlinespace[0ex]
    & & \multicolumn{2}{C{4.4cm}}{ Word2vec technique to compute word embeddings. } 
        & \multicolumn{2}{C{3.8cm}}{DNN}
        &  
        \begin{tabular}[t]{p{0.2cm}<{\centering}p{0.1cm}<{\centering}p{0cm}p{0.2cm}<{\centering}p{0.1cm}<{\centering}}
        ACC:&98.1\%&&PRE:&0.99\\
        REC:&0.96&&$F_{1}$:&0.97\\
        FPR:&0.01&&FNR:&0.04\\
        \end{tabular} \\                                        

        \cmidrule[0.8pt]{1-8}
        {\multirow{5}{1.5cm}{Achieving Control Flow Integrity \cite{yagemann2019barnum,phan2017convolutional,zhang2019deepcheck}}} 
                & {\multirow{2}{*}{Barnum\cite{yagemann2019barnum}}} & \multicolumn{3}{C{5cm}}{Phase I} & \multicolumn{3}{C{5cm}}{Phase II}  \\
        \cmidrule[\lightrulewidth](lr){3-8}\addlinespace[0ex]
        && \multicolumn{3}{C{5.7cm}}{The raw data, which is the exact sequence of instructions executed, was generated by combining the program binary, get immediately before the program opens a document, and Intel\textsuperscript{\textregistered} PT  trace. While Intel\textsuperscript{\textregistered} PT  built-in filtering options are set to CR3 and current privilege level (CPL), which only traces the program activity in the user space.  } 
            & \multicolumn{3}{C{5.7cm}}{The raw instruction sequences are summarized into Basic Blocks with IDs assigned and are then sliced into manageable subsequences with a fix window size 32, founded experimentally. Only sequences ending on indirect calls, jumps and returns are analyzed, since control-flow hijacking attacks always occur there. The label is the next BBID in the sequence.  } \\
        \cmidrule[\lightrulewidth](lr){3-8}\addlinespace[0ex]
        
        & & \multicolumn{2}{C{4.4cm}}{Phase III} & \multicolumn{2}{C{3.8cm}}{Phase IV} & \multicolumn{2}{C{2.8cm}}{Evaluation}  \\
        \cmidrule[\lightrulewidth](lr){3-8}\addlinespace[0ex]
        & & \multicolumn{2}{C{4.4cm}}{ N/A } 
            & \multicolumn{2}{C{3.8cm}}{LSTM}
            &  
            \begin{tabular}[t]{p{0.2cm}<{\centering}p{0.1cm}<{\centering}p{0cm}p{0.2cm}<{\centering}p{0.1cm}<{\centering}}
            ACC:&N/A\%&&PRE:&0.98\\
            REC:&1.00&&$F_{1}$:&0.98\\
            FPR:&0.98&&FNR:&0.02\\
            \end{tabular} \\  
        \cmidrule[1pt]{2-8}
        & {\multirow{2}{*}{CFG-CNN \cite{phan2017convolutional}}} & \multicolumn{3}{C{5cm}}{Phase I} & \multicolumn{3}{C{5cm}}{Phase II}  \\
        \cmidrule[\lightrulewidth](lr){3-8}\addlinespace[0ex]
        & & \multicolumn{3}{C{5.7cm}}{The raw data is instruction level control-flow graph constructed from program assembly code by an algorithm proposed by the authors. While in the CFG, one vertex corresponds to one instruction and one directed edge corresponds to an execution path from one instruction to another. The program sets for experiments are obtained from popular programming contest CodeChief.
        } 
            & \multicolumn{3}{C{5.7cm}}{Since each vertex of the CFG represents an instruction with complex information that could be viewed from different aspects, including instruction name, type, operands etc., a vertex is represented as the sum of a set of real valued vectors, corresponding to the number of views (e.g. \textit{addq 32,\%rsp} is converted to linear combination of randomly assigned vectors of \textit{addq value, reg}). The CFG is then sliced by a set of fixed size windows sliding through the entire graph to extract local features on different levels.} \\
        \cmidrule[\lightrulewidth](lr){3-8}\addlinespace[0ex]
        & & \multicolumn{2}{C{4.4cm}}{Phase III} & \multicolumn{2}{C{3.8cm}}{Phase IV} & \multicolumn{2}{C{2.8cm}}{Evaluation}  \\
        \cmidrule[\lightrulewidth](lr){3-8}\addlinespace[0ex]
        & & \multicolumn{2}{C{4.4cm}}{ N/A } 
            & \multicolumn{2}{C{3.8cm}}{DGCNN with different numbers of views and with or without operands}
            &  
            \begin{tabular}[t]{p{0.2cm}<{\centering}p{0.1cm}<{\centering}p{0cm}p{0.2cm}<{\centering}p{0.1cm}<{\centering}}
            ACC:&84.1\%&&PRE:&N/A\\
            REC:&N/A&&$F_{1}$:&N/A\\
            FPR:&N/A&&FNR:&N/A\\
            \end{tabular} \\                                                                                  
   \cmidrule[0.8pt]{1-8}
    {\multirow{5}{1.5cm}{Defending Network Attacks \cite{millar2018deep,zhang2019pccn,yuan2017deepdefense,varenne2019intelligent,yin2017deep,ustebay2019cyber,faker2019intrusion}}} 
            & {\multirow{2}{*}{50b(yte)-CNN\cite{millar2018deep}}} & \multicolumn{3}{C{5cm}}{Phase I} & \multicolumn{3}{C{5cm}}{Phase II}  \\
    \cmidrule[\lightrulewidth](lr){3-8}\addlinespace[0ex]
    && \multicolumn{3}{C{5.7cm}}{Open dataset UNSW-NB15 is used. First, tcpdump tool is utilised to capture 100 GB of the raw traffic (i.e. PCAP files) containing benign activities and 9 types of attacks. The Argus, Bro-IDS (now called Zeek) analysis tools are then used and twelve algorithms are developed to generate totally 49 features with the class label. In the end, the total number of data samples is 2,540,044 which are stored in CSV files.  } 
     & \multicolumn{3}{C{5.7cm}}{The first 50 bytes of each network traffic flow are picked out and each is directly used as one feature input to the neural network. } \\
    \cmidrule[\lightrulewidth](lr){3-8}\addlinespace[0ex]
    & & \multicolumn{2}{C{4.4cm}}{Phase III} & \multicolumn{2}{C{3.8cm}}{Phase IV} & \multicolumn{2}{C{2.8cm}}{Evaluation}  \\
    \cmidrule[\lightrulewidth](lr){3-8}\addlinespace[0ex]
    & & \multicolumn{2}{C{4.4cm}}{ N/A } 
      & \multicolumn{2}{C{3.8cm}}{ CNN with 2 hidden fully connected layers}
      &  
      \begin{tabular}[t]{p{0.2cm}<{\centering}p{0.1cm}<{\centering}p{0cm}p{0.2cm}<{\centering}p{0.1cm}<{\centering}}
      ACC:&N/A\%&&PRE:&N/A\\
      REC:&N/A&&$F_{1}$:&0.93\\
      FPR:&N/A&&FNR:&N/A\\
      \end{tabular} \\    
    \cmidrule[1pt]{2-8}
    & {\multirow{2}{*}{PCCN\cite{zhang2019pccn}}} & \multicolumn{3}{C{5cm}}{Phase I} & \multicolumn{3}{C{5cm}}{Phase II}  \\
    \cmidrule[\lightrulewidth](lr){3-8}\addlinespace[0ex]
    & & \multicolumn{3}{C{5.7cm}}{Open dataset CICIDS2017, which contains benign and 14 types of attacks, is used. Background benign network traffics are generated by profiling the abstract behavior of human interactions. Raw data are provided as PCAP files, and the results of the network traffic analysis using CICFlowMeter are pvodided as CSV files. In the end the dataset contains 3,119,345 data samples and 83 features categorized into 15 classes (1 normal + 14 attacks).   } 
     & \multicolumn{3}{C{5.7cm}}{Extract a total of 1,168,671 flow data, including 12 types of attack activities, from original dataset. Those flow data are then processed and visualized into grey-scale 2D graphs. The visualization method is not specified.} \\
    \cmidrule[\lightrulewidth](lr){3-8}\addlinespace[0ex]
    & & \multicolumn{2}{C{4.4cm}}{Phase III} & \multicolumn{2}{C{3.8cm}}{Phase IV} & \multicolumn{2}{C{2.8cm}}{Evaluation}  \\
    \cmidrule[\lightrulewidth](lr){3-8}\addlinespace[0ex]
    & & \multicolumn{2}{C{4.4cm}}{ N/A } 
      & \multicolumn{2}{C{3.8cm}}{Parallel cross CNN.}
      &  
      \begin{tabular}[t]{p{0.2cm}<{\centering}p{0.1cm}<{\centering}p{0cm}p{0.2cm}<{\centering}p{0.1cm}<{\centering}}
      ACC:&N/A\%&&PRE:&0.99\\
      REC:&N/A&&$F_{1}$:&0.99\\
      FPR:&N/A&&FNR:&N/A\\
      \end{tabular} \\    
    \cmidrule[0.8pt]{1-8}
    {\multirow{5}{1.5cm}{Malware Classification \cite{DeLaRosa2018,saxe2015deep,Kolosnjaji2017,McLaughlin2017,Tobiyama2016,Dahl2013,Nix2017,Kalash2018,Cui2018,David2015,Rosenberg2018,Xu2018}}} 
            & {\multirow{2}{*}{Rosenberg\cite{Rosenberg2018}}} & \multicolumn{3}{C{5cm}}{Phase I} & \multicolumn{3}{C{5cm}}{Phase II}  \\
    \cmidrule[\lightrulewidth](lr){3-8}\addlinespace[0ex]
    && \multicolumn{3}{C{5.7cm}}{The android dataset has the latest malware families and their variants, each with the same number of samples. The samples are labeled by VirusTotal. Then Cuckoo Sandbox is used to extract dynamic features (API calls) and static features (string). To avoid some anti-forensic sample, they applied YARA rule and removed sequences with less than 15 API calls. After preprocessing and balance the benign samples number, the dataset has 400,000 valid samples. } 
        & \multicolumn{3}{C{5.7cm}}{Long sequences cause out of memory during training LSTM model. So they use sliding window with fixed size and pad shorter sequences with zeros. One-hot encoding is applied to API calls. For static features strings, they defined a vector of 20,000 Boolean values indicating the most frequent Strings in the entire dataset. If the sample contain one string, the corresponding value in the vector will be assigned as 1, otherwise, 0.} \\
    \cmidrule[\lightrulewidth](lr){3-8}\addlinespace[0ex]
    
    & & \multicolumn{2}{C{4.4cm}}{Phase III} & \multicolumn{2}{C{3.8cm}}{Phase IV} & \multicolumn{2}{C{2.8cm}}{Evaluation}  \\
    \cmidrule[\lightrulewidth](lr){3-8}\addlinespace[0ex]
    & & \multicolumn{2}{C{4.4cm}}{ N/A } 
        & \multicolumn{2}{C{3.8cm}}{ They used RNN, BRNN, LSTM, Deep LSTM, BLSTM, Deep BLSTM, GRU, bi-directional GRU, Fully-connected DNN, 1D CNN in their experiments}
        &  
        \begin{tabular}[t]{p{0.2cm}<{\centering}p{0.1cm}<{\centering}p{0cm}p{0.2cm}<{\centering}p{0.1cm}<{\centering}}
        ACC:&98.3\%&&PRE:&N/A\\
        REC:&N/A&&$F_{1}$:&N/A\\
        FPR:&N/A&&FNR:&N/A\\
        \end{tabular} \\    

    \cmidrule[1pt]{2-8}
    & {\multirow{2}{*}{DeLaRosa\cite{DeLaRosa2018}}} & \multicolumn{3}{C{5cm}}{Phase I} & \multicolumn{3}{C{5cm}}{Phase II}  \\
    \cmidrule[\lightrulewidth](lr){3-8}\addlinespace[0ex]
    & & \multicolumn{3}{C{5.7cm}}{The windows dataset is from Reversing Labs including XP, 7, 8, and 10 for both 32-bit and 64-bit architectures and gathered over a span of twelve years (2006-2018). They selected nine malware families in their dataset and extracted static features in terms of bytes, basic, and assembly features.  } 
        & \multicolumn{3}{C{5.7cm}}{For bytes-level features, they used a sliding window to get the histogram of the bytes and compute the associated entropy in a window; for basic features, they created a fixed-sized feature vector given either a list of ASCII strings, or extracted import and metadata information from the PE Header(Strings are hashed and calculate a histogram of these hashes by counting the occurrences of each value); for assembly features, the disassembled code generated by Radare2 can be parsed and transformed into graph-like data structures such as call graphs, control flow graph, and instruction flow graph. } \\
    \cmidrule[\lightrulewidth](lr){3-8}\addlinespace[0ex]
    & & \multicolumn{2}{C{4.4cm}}{Phase III} & \multicolumn{2}{C{3.8cm}}{Phase IV} & \multicolumn{2}{C{2.8cm}}{Evaluation} \\
    \cmidrule[\lightrulewidth](lr){3-8}\addlinespace[0ex]
    & & \multicolumn{2}{C{4.4cm}}{ N/A} 
        & \multicolumn{2}{C{3.8cm}}{N/A}
        &  
        \begin{tabular}[t]{p{0.2cm}<{\centering}p{0.1cm}<{\centering}p{0cm}p{0.2cm}<{\centering}p{0.1cm}<{\centering}}
        ACC:&90.1\%&&PRE:&N/A\\
        REC:&N/A&&$F_{1}$:&N/A\\
        FPR:&N/A&&FNR:&N/A\\
        \end{tabular} \\    
    \cmidrule[0.8pt]{1-8}
    {\multirow{5}{1.5cm}{System Event Based Anomaly Detection\\\cite{Du:2017:deeLog,loganomaly:2019,Das:2018,Brown:2018,Zhang:2019,Bertero:8109100}}} 
            & {\multirow{2}{*}{DeepLog\cite{Du:2017:deeLog} }} & \multicolumn{3}{C{5cm}}{Phase I} & \multicolumn{3}{C{5cm}}{Phase II}  \\
    \cmidrule[\lightrulewidth](lr){3-8}\addlinespace[0ex]
    && \multicolumn{3}{C{5.7cm}}{More than 24 million raw log entries with the size of 2412 MB are recorded from the 203-node HDFS. Over 11 million log entries with 29 types are parsed, which are further grouped to 575,061 sessions according to block identifier. These sessions are manually labeled as normal and abnormal by HDFS experts. Finally, the constructed dataset HDFS 575,061 sessions of logs in the dataset, among which 16,838 sessions were labeled as anomalous } 
        & \multicolumn{3}{C{5.7cm}}{The raw log entries are parsed to different log type using Spell\cite{spell2016} which is based a longest common subsequence. There are total 29 log types in HDFS dataset} \\
    \cmidrule[\lightrulewidth](lr){3-8}\addlinespace[0ex]
    
    & & \multicolumn{2}{C{4.4cm}}{Phase III} & \multicolumn{2}{C{3.8cm}}{Phase IV} & \multicolumn{2}{C{2.8cm}}{Evaluation}  \\
    \cmidrule[\lightrulewidth](lr){3-8}\addlinespace[0ex]
    & & \multicolumn{2}{C{4.4cm}}{DeepLog directly utilized one-hot vector to represent 29 log key without represent learning} 
        & \multicolumn{2}{C{3.8cm}}{ A stacked LSTM with two hidden LSTM layers.}
        &  
        \begin{tabular}[t]{p{0.2cm}<{\centering}p{0.1cm}<{\centering}p{0cm}p{0.2cm}<{\centering}p{0.1cm}<{\centering}}
        ACC:&N/A\%&&PRE:&0.95\\
        REC:&0.96&&$F_{1}$:&0.96\\
        FPR:&N/A&&FNR:&N/A\\
        \end{tabular} \\    
    \cmidrule[1pt]{2-8}
    & {\multirow{2}{*}{LogAnom \cite{loganomaly:2019}}} & \multicolumn{3}{C{5cm}}{Phase I} & \multicolumn{3}{C{5cm}}{Phase II}  \\
    \cmidrule[\lightrulewidth](lr){3-8}\addlinespace[0ex]
    & & \multicolumn{3}{C{5.7cm}}{LogAnom also used HDFS dataset, which is same as DeepLog. } 
        & \multicolumn{3}{C{5.7cm}}{The raw log entries are parsed to different log templates using FT-Tree \cite{ftree2017} according the frequent combinations of log words. There are total 29 log templates in HDFS dataset} \\
    \cmidrule[\lightrulewidth](lr){3-8}\addlinespace[0ex]
    & & \multicolumn{2}{C{4.4cm}}{Phase III} & \multicolumn{2}{C{3.8cm}}{Phase IV} & \multicolumn{2}{C{2.8cm}}{Evaluation}  \\
    \cmidrule[\lightrulewidth](lr){3-8}\addlinespace[0ex]
    & & \multicolumn{2}{C{4.4cm}}{ LogAnom employed Word2Vec to represent the extracted log templates with more semantic information } 
        & \multicolumn{2}{C{3.8cm}}{Two LSTM  layers with 128 neurons}
        &  
        \begin{tabular}[t]{p{0.2cm}<{\centering}p{0.1cm}<{\centering}p{0cm}p{0.2cm}<{\centering}p{0.1cm}<{\centering}}
        ACC:&N/A\%&&PRE:&0.97\\
        REC:&0.94&&$F_{1}$:&0.96\\
        FPR:&N/A&&FNR:&N/A\\
        \end{tabular} \\       
    \cmidrule[0.8pt]{1-8}
    {\multirow{5}{1.5cm}{Memory Forensics \cite{Song:2018:DeepMem,Petrik:2018,Michalas:2017,DAI201830}}} 
            & {\multirow{2}{*}{DeepMem\cite{Song:2018:DeepMem}}} & \multicolumn{3}{C{5cm}}{Phase I} & \multicolumn{3}{C{5cm}}{Phase II}  \\
    \cmidrule[\lightrulewidth](lr){3-8}\addlinespace[0ex]
    && \multicolumn{3}{C{5.7cm}}{400 memory dumps are collected on Windows 7 x86 SP1 virtual machine with simulating various random user actions and forcing the OS to randomly allocate objects. The size of each dump is 1GB. } 
        & \multicolumn{3}{C{5.7cm}}{Construct memory graph from memory dumps, where each node represents a segment between two pointers and an edge is created if two nodes are neighbor} \\
    \cmidrule[\lightrulewidth](lr){3-8}\addlinespace[0ex]
    
    & & \multicolumn{2}{C{4.4cm}}{Phase III} & \multicolumn{2}{C{3.8cm}}{Phase IV} & \multicolumn{2}{C{2.8cm}}{Evaluation}  \\
    \cmidrule[\lightrulewidth](lr){3-8}\addlinespace[0ex]
    & & \multicolumn{2}{C{4.4cm}}{Each node is represented by a latent numeric vector from the embedding network.} 
        & \multicolumn{2}{C{3.8cm}}{ Fully Connected Network (FCN) with ReLU layer.}
        &  
        \begin{tabular}[t]{p{0.2cm}<{\centering}p{0.1cm}<{\centering}p{0cm}p{0.2cm}<{\centering}p{0.1cm}<{\centering}}
        ACC:&N/A\%&&PRE:&0.99\\
        REC:&0.99&&$F_{1}$:&0.99\\
        FPR:&0.01&&FNR:&0.01\\
        \end{tabular} \\    
    \cmidrule[1pt]{2-8}
    & {\multirow{2}{*}{MDMF~\cite{Petrik:2018}}} & \multicolumn{3}{C{5cm}}{Phase I} & \multicolumn{3}{C{5cm}}{Phase II}  \\
    \cmidrule[\lightrulewidth](lr){3-8}\addlinespace[0ex]
    & & \multicolumn{3}{C{5.7cm}}{Create a dataset of benign host memory snapshots running normal, non-compromised software, including software that executes in many of the malicious snapshots. The benign snapshot is extracted from memory after ample time has passed for the chosen programs to open. By generating samples in parallel to the separate malicious environment, the benign memory snapshot dataset created. } 
        & \multicolumn{3}{C{5.7cm}}{Various representation for the memory snapshots including byte sequence and image, without relying on domain-knowledge of the OS.} \\
    \cmidrule[\lightrulewidth](lr){3-8}\addlinespace[0ex]
    & & \multicolumn{2}{C{4.4cm}}{Phase III} & \multicolumn{2}{C{3.8cm}}{Phase IV} & \multicolumn{2}{C{2.8cm}}{Evaluation}  \\
    \cmidrule[\lightrulewidth](lr){3-8}\addlinespace[0ex]
    & & \multicolumn{2}{C{4.4cm}}{ N/A} 
        & \multicolumn{2}{C{3.8cm}}{Recurrent Neural Network with LSTM cells and Convolutional Neural Network composed of multiple layers, including pooling and fully connected layers. for image data}
        &  
        \begin{tabular}[t]{p{0.2cm}<{\centering}p{0.1cm}<{\centering}p{0cm}p{0.2cm}<{\centering}p{0.1cm}<{\centering}}
        ACC:&98.0\%&&PRE:&N/A\\
        REC:&N/A&&$F_{1}$:&N/A\\
        FPR:&N/A&&FNR:&N/A\\
        \end{tabular} \\    
    \cmidrule[0.8pt]{1-8}
    {\multirow{5}{1.5cm}{Fuzzing \cite{wang2019neufuzz,shi2019neuzz,bottinger2018deep,godefroid2017learn,rajpal2017not}}} 
            & {\multirow{2}{*}{L-Fuzz\cite{godefroid2017learn}}} & \multicolumn{3}{C{5cm}}{Phase I} & \multicolumn{3}{C{5cm}}{Phase II}  \\
    \cmidrule[\lightrulewidth](lr){3-8}\addlinespace[0ex]
    && \multicolumn{3}{C{5.7cm}}{The raw data are about 63,000 non-binary PDF objects, sliced in fix size, extracted from 534 PDF files that are provided by Windows fuzzing team and are previously used for prior extended fuzzing of Edge PDF parser.  } 
        & \multicolumn{3}{C{5.7cm}}{N/A} \\
    \cmidrule[\lightrulewidth](lr){3-8}\addlinespace[0ex]
    
    & & \multicolumn{2}{C{4.4cm}}{Phase III} & \multicolumn{2}{C{3.8cm}}{Phase IV} & \multicolumn{2}{C{2.8cm}}{Evaluation}  \\
    \cmidrule[\lightrulewidth](lr){3-8}\addlinespace[0ex]
    & & \multicolumn{2}{C{4.4cm}}{N/A} 
        & \multicolumn{2}{C{3.8cm}}{ Char-RNN}
        &  
        \begin{tabular}[t]{p{0.2cm}<{\centering}p{0.1cm}<{\centering}p{0cm}p{0.2cm}<{\centering}p{0.1cm}<{\centering}}
        ACC:&N/A\%&&PRE:&N/A\\
        REC:&N/A&&$F_{1}$:&0.93\\
        FPR:&N/A&&FNR:&N/A\\
        \end{tabular} \\    
    \cmidrule[1pt]{2-8}
    & {\multirow{2}{*}{NEUZZ\cite{shi2019neuzz} }} & \multicolumn{3}{C{5cm}}{Phase I} & \multicolumn{3}{C{5cm}}{Phase II}  \\
    \cmidrule[\lightrulewidth](lr){3-8}\addlinespace[0ex]
    & & \multicolumn{3}{C{5.7cm}}{For each program tested, the raw data is collected by running AFL-2.52b on a single core machine for one hour. The training data are byte level input files generated by AFL, and the labels are bitmaps corresponding to input files. For experiments, NEUZZ is implemented on 10 real-world programs, the LAVA-M bug dataset, and the CGC dataset.
    } 
        & \multicolumn{3}{C{5.7cm}}{N/A } \\
    \cmidrule[\lightrulewidth](lr){3-8}\addlinespace[0ex]
    & & \multicolumn{2}{C{4.4cm}}{Phase III} & \multicolumn{2}{C{3.8cm}}{Phase IV} & \multicolumn{2}{C{2.8cm}}{Evaluation}  \\
    \cmidrule[\lightrulewidth](lr){3-8}\addlinespace[0ex]
    & & \multicolumn{2}{C{4.4cm}}{ N/A} 
        & \multicolumn{2}{C{3.8cm}}{NN }
        &  
        \begin{tabular}[t]{p{0.2cm}<{\centering}p{0.1cm}<{\centering}p{0cm}p{0.2cm}<{\centering}p{0.1cm}<{\centering}}
        ACC:&N/A\%&&PRE:&N/A\\
        REC:&N/A&&$F_{1}$:&0.93\\
        FPR:&N/A&&FNR:&N/A\\
        \end{tabular} \\                                                                                                                          
    \bottomrule
    \bottomrule
    \end{longtable}
    \begin{TableNotes}
        \item [1] Deep Learning metrics are often not available in fuzzing papers. Typical fuzzing metrics used for evaluations are: code coverage, pass rate and bugs.
    \end{TableNotes}
\end{ThreePartTable}
\end{center}

\subsection{Methodology for reviewing the existing works} 
\label{lab:three_questions}

Data representation (or feature engineering) plays an important role in solving security problems with Deep Learning. This is because data representation is a way to take advantage of human ingenuity and prior knowledge to extract and organize the discriminative information from the data. Many efforts in deploying machine learning algorithms in security domain actually goes into the design of preprocessing pipelines and data transformations that result in a representation of the data to support effective machine learning. 

In order to expand the scope and ease of applicability of machine learning in security domain, it would be highly desirable to find a proper way to represent the data in security domain, which can entangle and hide more or less the different explanatory factors of variation behind the data. 
To let this survey adequately reflect the important role
played by data representation, our review will focus on how the following three questions are answered by the existing works:
\label{threequestions}
\begin{itemize}
\item {\bf Question 1:} Is Phase~\ref{phase2} pervasively done in the literature? When Phase~\ref{phase2} is skipped in a work, are there any particular reasons?  

\item {\bf Question 2:} Is Phase~\ref{phase3} employed in the literature? 
 When Phase~\ref{phase3} is skipped in a work, are there any particular reasons? 

\item {\bf Question 3:} When solving different security problems, is there any commonality in terms of the (types of) classifiers learned in Phase~\ref{phase4}? Among the works solving the same security problem, is there dissimilarity in terms of classifiers learned in Phase~\ref{phase4}?

\end{itemize}

\newlength\treeheight
\setlength{\treeheight}{8cm}
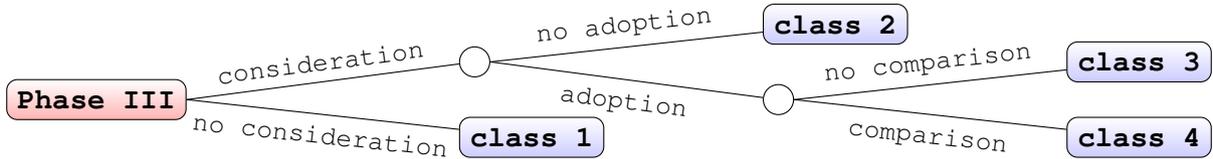
\begin{figure}[htbp]
  \centering 
\begin{tikzpicture}[
  grow                    = right,
  sibling distance        = 5em,
  anchor=west,
     growth parent anchor=east, 
     parent anchor=east, 
     level distance=0.5cm, 
  every node/.style={},
  level 1/.style={sibling distance=\treeheight/8},
  level 2/.style={sibling distance=\treeheight/8},
  level distance          = 8.5em,
  every node/.style       = {font=\footnotesize},
  sloped
]
  \node [root] {\textbf{Phase~\ref{phase3}}}
  child { node [env] {\textbf{class 1}}
      edge from parent node [below] {\texttt{no consideration}} }
    child { node [dummy] {}
        child { node [dummy] {}
            child {node(pass) [env] {\textbf{class 4}}
                edge from parent node [below, align=center]{\texttt{comparison}}}
            child {node [env] {\textbf{class 3}}
                edge from parent node [above, align=center]{\texttt{no comparison}}}
            edge from parent node [below] {\texttt{adoption}} }
        child { node(plugin) [env] {\textbf{class 2}}
            edge from parent node [above, align=center]{\texttt{no adoption}}}
        edge from parent node [above] {\texttt{consideration}}};
\end{tikzpicture}
\caption{Classification tree for different Phase~\ref{phase3} methods. Here, \texttt{consideration}, \texttt{adoption}, and \texttt{comparison} indicate that a work considers Phase~\ref{phase3}, adopts Phase~\ref{phase3} and makes comparison with other methods, respectively.} 
\label{fig:dec}
\end{figure}

To group the Phase~\ref{phase3} methods at different applications of Deep Learning in solving the same security problem, we introduce a classification tree as shown in Figure~\ref{fig:dec}. The classification tree categorizes the Phase~\ref{phase3} methods in our selected survey works into four classes. First, class 1 includes the Phase~\ref{phase3} methods which do not consider representation learning. Second, class 2 includes the Phase~\ref{phase3} methods which consider representation learning but, do not adopt it. Third, class 3 includes the Phase~\ref{phase3} methods which consider and adopt representation learning but, do not compare the performance with other methods. Finally, class 4 includes the Phase~\ref{phase3} methods which consider and adopt representation learning and, compare the performance with other methods.

In the remaining of this paper, we take a closer look at how each of the eight security problems is being solved by applications of Deep Learning in the literature.

\section{A closer look at applications of Deep Learning in solving security-oriented program analysis challenges}
\label{sec:programanalysis}

\subsection{Introduction}
Recent years, security-oriented program analysis is widely used in software security. For example, symbolic execution and taint analysis are used to discover, detect and analyze vulnerabilities in programs. Control flow analysis, data flow analysis and pointer/alias analysis are important components 
when enforcing many secure strategies, 
such as control flow integrity, data flow integrity and doling dangling pointer elimination.
Reverse engineering was used by defenders and attackers to understand the logic of a program without source code.

{\color{myblack} In the security-oriented program analysis, there are many open problems, such as precise pointer/alias analysis, 
accurate and complete reversing engineer, complex constraint solving, program de-obfuscation, and so on. 
Some problems have theoretically proven to be NP-hard, and others still need lots of human effort to solve. 
Either of them needs a lot of domain knowledge and experience from expert to develop better solutions. 
Essentially speaking, the main challenges when solving them through traditional approaches 
are due to the sophisticated rules between the features and labels, which may change in different contexts. 
Therefore, on the one hand, it will take a large quantity of human effort to develop rules to solve the problems, 
on the other hand, even the most experienced expert cannot guarantee completeness. 
Fortunately, the deep learning method is skillful to find relations between features and labels if given a large amount of training data. 
It can quickly and comprehensively find all the relations if the training samples are representative and effectively encoded.}

In this section, we will review the very recent four representative works that use Deep Learning for security-oriented program analysis. We observed that they focused on different goals. Shin, et al. designed a model~\cite{shin2015recognizing} to identify the function boundary. EKLAVYA~\cite{chua2017neural} was developed to learn the function type. Gemini~\cite{xu2017neural} was proposed to detect similarity among functions.  DEEPVSA~\cite{guo2019deepvsa} was designed to learn memory region of an indirect addressing from the code sequence. Among these works, we select two representative works~\cite{shin2015recognizing,chua2017neural} and then, 
summarize the analysis results in Table~\ref{Table:Summary} in detail.

Our review will be centered around three questions described in Section~\ref{threequestions}. In the remaining of this section, we will first provide a set of observations, and then we provide the indications. Finally, we provide some general remarks.

\subsection{Key findings from a closer look}
From a close look at the very recent applications using Deep Learning for solving security-oriented program analysis challenges, we observed the followings: 

\begin{itemize}[label={}]
\item  \observation{} \textit{All of the works in our survey used binary files as their raw data.}

Phase~\ref{phase2} in our survey had one similar and straightforward goal~\textendash~extracting code sequences from the binary. Difference among them was that the code sequence was extracted directly from the binary file when solving problems in static program analysis, 
while it was extracted from the program execution when solving problems in dynamic program analysis.

\item  \observation{*} \textit{Most data representation methods generally took into account the domain knowledge.}

Most data representation methods generally took into the domain knowledge, i.e., what kind of information they wanted to reserve when processing their data. Note that the feature selection has a wide influence on Phase~\ref{phase2} and Phase~\ref{phase3}, for example, embedding granularities, representation learning methods. Gemini~\cite{xu2017neural} selected function level feature and other works in our survey selected instruction level feature. To be specifically, all the works except Gemini~\cite{xu2017neural} vectorized code sequence on instruction level.

\item  \observation{} 
\textit{To better support data representation for high performance, some works adopted representation learning.}

For instance, DEEPVSA~\cite{guo2019deepvsa} employed a representation learning method, i.e., bi-directional LSTM, to learn data dependency within instructions. EKLAVYA~\cite{chua2017neural} adopted representation learning method, i.e., word2vec technique, to extract inter-instruciton information. It is worth noting that Gemini~\cite{xu2017neural} adopts the Structure2vec embedding network in its siamese architecture in Phase~\ref{phase4} (see details in Observation~3.\ref{obs:bi}). The Structure2vec embedding network learned information from an attributed control flow graph.

\item  \observation{} \textit{According to our taxonomy, most works in our survey were classified into class 4.}

To compare the Phase~\ref{phase3}, we introduced a 
classification tree with three layers as shown in Figure~\ref{fig:dec} to group different works into four categories. The decision tree grouped our surveyed works into four classes according to whether they considered representation learning or not, whether they adopted representation learning or not, and whether they compared their methods with others', respectively, when designing their framework. According to our taxonomy, EKLAVYA~\cite{chua2017neural}, DEEPVSA~\cite{guo2019deepvsa} were grouped into class 4 shown in Figure~\ref{fig:dec}. Also, Gemini's work~\cite{xu2017neural} and Shin, et al.'s work~\cite{shin2015recognizing} belonged to class 1 and class 2 shown in Figure~\ref{fig:dec}, respectively.

\item  \observation{} \label{obs:reason} \textit{All the works in our survey explain why they adopted or did not adopt one of representation learning algorithms.}

Two works in our survey adopted representation learning for different reasons: to enhance model's ability of generalization~\cite{chua2017neural}; and to learn the dependency within instructions~\cite{guo2019deepvsa}.
It is worth noting that Shin, et al. did not adopt representation learning because they wanted to preserve the ``attractive'' features of neural networks over other machine learning methods~\textendash~simplicity. As they stated, ``first, neural networks can learn directly from the original representation with minimal preprocessing (or ``feature engineering”) needed." and ``second, neural networks can learn end-to-end, where each of its constituent stages are trained simultaneously in order to best solve the end goal." Although Gemini~\cite{xu2017neural} did not adopt representation learning when processing their raw data, the Deep Learning models in siamese structure consisted of two graph embedding networks and one cosine function.


\item  \observation{*} \textit{The analysis results showed that a suitable representation learning method could improve accuracy of Deep Learning models.}

DEEPVSA~\cite{guo2019deepvsa} designed a series of experiments to evaluate the effectiveness of its representative method. By combining with the domain knowledge, EKLAVYA~\cite{chua2017neural} employed t-SNE plots and analogical reasoning to explain the effectiveness of their representation learning method in an intuitive way.

\item  \observation{*} \label{obs:bi} \textit{Various Phase IV methods were used.}

In Phase~\ref{phase4}, Gemini~\cite{xu2017neural} adopted siamese architecture model which consisted of two Structure2vec embedding networks and one cosine function. 
The siamese architecture took two functions as its input, and produced the similarity score as the output. The other three works~\cite{shin2015recognizing,chua2017neural,guo2019deepvsa} adopted bi-directional RNN, RNN, bi-directional LSTM respectively. Shin, et al. adopted bi-directional RNN because they wanted to combine both the past and the future information in making a prediction for the present instruction~\cite{shin2015recognizing}. DEEPVSA~\cite{guo2019deepvsa} adopted bi-directional RNN to enable their model to infer memory regions in both forward and backward ways. 
\end{itemize} 
 
The above observations seem to indicate the following indications: 
\begin{itemize}[label={}]
\item \indication{} \textit{Phase~\ref{phase3} is not always necessary.}

Not all authors regard representation learning as a good choice
even though some case experiments show that representation learning can improve the final results.
They value more the simplicity of Deep Learning methods and suppose that the adoption of representation learning weakens the simplicity of Deep Learning methods.

\item \indication{}
\textit{Even though the ultimate objective of Phase~\ref{phase3} in the four surveyed works is to train a model with better accuracy, 
they have different specific motivations as described in Observation~3.\ref{obs:reason}.}

When authors choose representation learning, 
they usually try to convince people the effectiveness of their choice by 
empirical or theoretical analysis. 

\item \indication{*} 
\textit{3.\ref{obs:bi} indicates that authors usually refer to the domain knowledge when designing the architecture of Deep Learning model.}

For instance, the works we reviewed commonly adopt bi-directional RNN when their prediction partly based on future information in data sequence.



\end{itemize} 
{\color{myblack}
\subsection{Discussion}
\label{sec:pa:dis}
Despite the effectiveness and agility of deep learning-based methods, 
there are still some challenges in developing a scheme with high accuracy due to the hierarchical data structure, 
lots of noisy, and unbalanced data composition in program analysis. For instance, an instruction sequence, a typical data sample in program analysis, 
contains three-level hierarchy: sequence--instruction--opcode/operand. 
To make things worse, each level may contain many different structures, e.g., one-operand instructions, multi-operand instructions,
which makes it harder to encode the training data.
}





\section{A closer look at applications of Deep Learning in defending ROP attacks}
\label{sec:rop}
\subsection{Introduction}
Return-oriented programming (ROP) attack is one of the most dangerous code reuse attacks, which allows the attackers to launch control-flow hijacking attack without injecting any malicious code. Rather, It leverages particular instruction sequences (called ``gadgets'') widely existing in the program space to achieve Turing-complete attacks~\cite{shacham2007geometry}. Gadgets are instruction sequences that end with a \textit{RET} instruction. 
Therefore, they can be chained together by specifying the return addresses on program stack.
Many traditional techniques could be used to detect ROP attacks, such as control-flow integrity (CFI~\cite{abadi2009control}), 
but many of them either have low detection rate or have high runtime overhead. 
ROP payloads do not contain any codes. 
In other words, analyzing ROP payload without the context of the program’s memory dump is meaningless. 
Thus, the most popular way of detecting and preventing ROP attacks is control-flow integrity. 
{\color{myblack} The challenge after acquiring the instruction sequences is that it is hard to recognize whether the control flow is normal. 
Traditional methods use the control flow graph (CFG) to identify whether the control flow is normal, 
but attackers can design the instruction sequences which follow the normal control flow defined by the CFG.
In essence, it is very hard to design a CFG to exclude every single possible combination of instructions that can be used to launch ROP attacks. 
Therefore, using data-driven methods could help eliminate such problems.}

In this section, we will review the very recent three representative works that use Deep Learning for defending ROP attacks: ROPNN \cite{li2018ropnn}, HeNet \cite{chen2018henet} and DeepCheck \cite{zhang2019deepcheck}. 
ROPNN \cite{li2018ropnn} aims to detect ROP attacks, HeNet \cite{chen2018henet} aims to detect malware using CFI, and DeepCheck \cite{zhang2019deepcheck} aims at detecting all kinds of code reuse attacks.

{\color{myblack} Specifically, ROPNN is to protect one single program at a time, and its training data are generated from real-world programs along with their execution. 
Firstly, it generates its benign and malicious data by ``chaining-up" the normally executed instruction sequences and
``chaining-up'' gadgets with the help of gadgets generation tool, respectively, 
after the memory dumps of programs are created.
Each data sample is byte-level instruction sequence labeled as ``benign'' or ``malicious''. 
Secondly, ROPNN will be trained using both malicious and benign data.
Thirdly, the trained model is deployed to a target machine. After the protected program started, 
the executed instruction sequences will be traced and fed into the trained model, 
the protected program will be terminated once the model found the instruction sequences are likely to be malicious.

HeNet is also proposed to protect a single program.
Its malicious data and benign data are generated by collecting trace data through Intel PT from malware and normal software, respectively.
Besides, HeNet preprocesses its dataset and shape each data sample in the format of image,
so that they could implement transfer learning from a model pre-trained on ImageNet. 
Then, HeNet is trained and deployed on machines with features of Intel PT to collect and classify the program's execution trace online. 

The training data for DeepCheck are acquired from CFGs, 
which are constructed by dissembling the programs and using the information from Intel PT. 
After the CFG for a protected program is constructed, 
authors sample benign instruction sequences by chaining up basic blocks that are connected by edges, 
and sample malicious instruction sequences by chaining up those that are not connected by edges.
Although a CFG is needed during training, there is no need to construct CFG after the training phase. 
After deployed, instruction sequences will be constructed by leveraging Intel PT on the protected program. 
Then the trained model will classify whether the instruction sequences are malicious or benign.} 

We observed that none of the works considered Phase~\ref{phase3}, so all of them belong to class 1 according to our taxonomy as shown in Figure~\ref{fig:dec}. The analysis results of ROPNN \cite{li2018ropnn} and HeNet \cite{chen2018henet} are shown in Table \ref{Table:Summary}. Also, we observed that three works had different goals. 

Our review will be centered around three questions described in Section~\ref{threequestions}. In the remaining of this section, we will first provide a set of observations, and then we provide the indications. Finally, we provide some general remarks.

\subsection{Key findings from a closer look}

From a close look at the very recent applications using Deep Learning for defending return-oriented programming attacks, we observed the followings: 

\begin{itemize}[label={}]

\item  \observation{}\label{rop:obs1} \textit{All the works\cite{li2018ropnn, zhang2019deepcheck,chen2018henet} in this survey focused on data generation and acquisition.}

In ROPNN \cite{li2018ropnn}, both malicious samples (gadget chains) were generated using an automated gadget generator (i.e. ROPGadget~\cite{ROPgadget}) and a CPU emulator (i.e. Unicorn~\cite{unicorn}). ROPGadget was used to extract instruction sequences that could be used as gadgets from a program, and Unicorn was used to validate the instruction sequences.
Corresponding benign sample (gadget-chain-like instruction sequences) were generated by disassembling a set of programs.
In DeepCheck~\cite{zhang2019deepcheck} refers to the key idea of control-flow integrity~\cite{abadi2009control}. It generates program's run-time control flow through new feature of Intel CPU (Intel Processor Tracing), then compares the run-time control flow with the program's control-flow graph (CFG) that generates through static analysis. Benign instruction sequences are that with in the program's CFG, and vice versa. In HeNet~\cite{chen2018henet}, program's execution trace was extracted using the similar way as DeepCheck. Then, each byte was transformed into a pixel with an intensity between 0-255. Known malware samples and benign software samples were used to generate malicious data benign data, respectively.

\item  \observation{}\label{rop:obs2} \textit{None of the ROP works in this survey deployed Phase~\ref{phase3}.}

Both ROPNN\cite{li2018ropnn} and DeepCheck\cite{zhang2019deepcheck} used binary instruction sequences for training. In ROPNN\cite{li2018ropnn}, one byte was used as the very basic element for data pre-processing. Bytes were formed into one-hot matrices and flattened for 1-dimensional convolutional layer. In DeepCheck \cite{zhang2019deepcheck}, half-byte was used as the basic unit. Each half-byte (4 bits) was transformed to decimal form ranging from 0-15 as the basic element of the input vector, then was fed into a fully-connected input layer. On the other hand, HeNet \cite{chen2018henet} used different kinds of data. By the time this survey has been drafted, the source code of HeNet was not available to public and thus, the details of the data pre-processing was not be investigated. However, it is still clear that HeNet used binary branch information collected from Intel PT rather than binary instructions. In HeNet, each byte was converted to one decimal number ranging from 0 to 255. Byte sequences was sliced and formed into image sequences (each pixel represented one byte) for a fully-connected input layer.

\item  \observation{}\label{rop:obs3} \textit{Fully-connected neural network was widely used.}

Only ROPNN \cite{li2018ropnn} used 1-dimensional convolutional neural network (CNN) when extracting features. Both HeNet \cite{chen2018henet} and DeepCheck\cite{zhang2019deepcheck} used fully-connected neural network (FCN). None of the works used recurrent neural network (RNN) and the variants.
\end{itemize}
 
The above observations seem to indicate the following indications:

\begin{itemize}

\item \indication{} \textit{It seems like that one of the most important factors in ROP problem is feature selection and data generation.} 

All three works use very different methods to collect/generate data, and all the authors provide very strong evidences and/or arguments to justify their approaches. ROPNN~\cite{li2018ropnn} was trained by the malicious and benign instruction sequences. However, there is no clear boundary between benign instruction sequences and malicious gadget chains. This weakness may impair the performance when applying ROPNN to real world ROP attacks. As oppose to ROPNN, DeepCheck \cite{zhang2019deepcheck} utilizes CFG to generate training basic-block sequences. However, since the malicious basic-block sequences are generated by randomly connecting nodes without edges, it is not guaranteed that all the malicious basic-blocks are executable. HeNet \cite{chen2018henet} generates their training data from malware. Technically, HeNet could be used to detect any binary exploits, but their experiment focuses on ROP attack and achieves 100\% accuracy. This shows that the source of data in ROP problem does not need to be related to ROP attacks to produce very impressive results.

\item \indication{} \textit{Representation learning seems not critical when solving ROP problems using Deep Learning.} 

Minimal process on data in binary form seems to be enough to transform the data into a representation that is suitable for neural networks. Certainly, it is also possible to represent the binary instructions at a higher level, such as opcodes, or use embedding learning. However, as stated in \cite{li2018ropnn}, it appears that the performance will not change much by doing so. The only benefit of representing input data to a higher level is to reduce irrelevant information, but it seems like neural network by itself is good enough at extracting features.

\item \indication{} \textit{Different Neural network architecture does not have much influence on the effectiveness of defending ROP attacks.}

Both HeNet \cite{chen2018henet} and DeepCheck \cite{zhang2019deepcheck} utilizes standard DNN and achieved comparable results on ROP problems. One can infer that the input data can be easily processed by neural networks, and the features can be easily detected after proper pre-process. 

\end{itemize} 

It is not surprising that researchers are not very interested in representation learning for ROP problems as stated in Observation~4.\ref{rop:obs1}. 
Since ROP attack is focus on the gadget chains, it is straightforward for the researcher to choose the gadgets as their training data directly.
It is easy to map the data into numerical representation with minimal processing. An example is that one can map binary executable to hexadecimal ASCII representation, which could be a good representation for neural network.

Instead, researchers focus more in data acquisition and generation. In ROP problems, the amount of data is very limited. Unlike malware and logs, ROP payloads normally only contain addresses rather than codes, which do not contain any information without providing the instructions in corresponding addresses. It is thus meaningless to collect all the payloads. At the best of our knowledge, all the previous works use pick instruction sequences rather than payloads as their training data, even though they are hard to collect.

{\color{myblack}
\subsection{Discussion}
Even though, Deep Learning based method does not face the challenge to design a very complex fine-grained CFG anymore,
it suffers from a limited number of data sources.
Generally, Deep Learning based method requires lots of training data.
However, real-world malicious data for the ROP attack is very hard to find, 
because comparing with benign data, malicious data need to be carefully crafted and there is no existing database to collect all the ROP attacks. 
Without enough representative training set, the accuracy of the trained model cannot be guaranteed.
}

\section{A closer look at applications of Deep Learning in achieving CFI}
\label{sec:cfi}
\subsection{Introduction}
The basic ideas of control-flow integrity (CFI) techniques, proposed by Abadi in 2005\cite{abadi2009control}, could be dated back to 2002, when Vladimir and his fellow researchers proposed an idea called program shepherding\cite{kiriansky2002secure}, a method of monitoring the execution flow of a program when it is running by enforcing some security policies. The goal of CFI is to detect and prevent control-flow hijacking attacks, by restricting every critical control flow transfers to a set that can only appear in correct program executions, according to a pre-built CFG. Traditional CFI techniques typically leverage some knowledge, gained from either dynamic or static analysis of the target program, combined with some code instrumentation methods, to ensure the program runs on a correct track. 

{\color{myblack}However, the problems of traditional CFI are: (1) Existing CFI implementations are not compatible with some of important code features~\cite{236352}; 
(2) CFGs generated by static, dynamic or combined analysis cannot always be precisely completed due to some open problems~\cite{pointernp}; 
(3) There always exist certain level of compromises between accuracy and performance overhead and other important properties~\cite{tan2017cfg,wang2019gpt}.}
Recent research has proposed to apply Deep Learning on detecting control flow violation. Their result shows that, compared with traditional CFI implementation, the security coverage and scalability were enhanced in such a fashion~\cite{yagemann2019barnum}. 
Therefore, we argue that Deep Learning could be another approach which requires more attention from CFI researchers who aim at achieving control-flow integrity more efficiently and accurately.

In this section, we will review the very recent three representative papers that use Deep Learning for achieving CFI. Among the three, two representative papers\cite{yagemann2019barnum,phan2017convolutional} are already summarized phase-by-phase in Table \ref{Table:Summary}. We refer to interested readers the Table~\ref{Table:Summary} for a concise overview of those two papers. 

Our review will be centered around three questions described in Section~\ref{threequestions}. In the remaining of this section, we will first provide a set of observations, and then we provide the indications. Finally, we provide some general remarks.

\subsection{Key findings from a closer look}

From a close look at the very recent applications using Deep Learning for achieving control-flow integrity, we observed the followings: 

\begin{itemize}[label={}]
\item  \observation{} \textit{None of the related works realize preventive\footnote{We refer readers to~\cite{wang2019gpt} which systemizes the knowledge of protections by CFI schemes.} prevention of control flow violation.}

After doing a thorough literature search, we observed that security researchers are quite behind the trend of applying Deep Learning techniques to solve security problems. Only one paper has been founded by us, using Deep Learning techniques to directly enhance the performance of CFI~\cite{yagemann2019barnum}. This paper leveraged Deep Learning to detect document malware through checking program's execution traces that generated by hardware. Specifically, the CFI violations were checked in an offline mode. So far, no works have realized Just-In-Time checking for program's control flow.

In order to provide more insightful results, in this section, we try not to narrow down our focus on CFI detecting attacks at run-time, but to extend our scope to papers that take good use of control flow related data, combined with Deep Learning techniques\cite{phan2017convolutional,nguyen2018auto}. In one work, researchers used self-constructed instruction-level CFG to detect program defection\cite{phan2017convolutional}. In another work, researchers used lazy-binding CFG to detect sophisticated malware~\cite{nguyen2018auto}. 
 
\item  \observation{} \textit{Diverse raw data were used for evaluating CFI solutions.}

In all surveyed papers, there are two kinds of control flow related data being used: program instruction sequences and CFGs. Barnum et al.~\cite{yagemann2019barnum} employed statically and dynamically generated instruction sequences acquired by program disassembling and Intel\textsuperscript{\textregistered} Processor Trace. CNNoverCFG~\cite{phan2017convolutional} used self-designed algorithm to construct instruction level control-flow graph. Minh Hai Nguyen et al.~\cite{nguyen2018auto} used proposed lazy-binding CFG to reflect the behavior of malware DEC.

\item  \observation{} \textit{All the papers in our survey adopted Phase~\ref{phase2}.}

All the related papers in our survey employed Phase~\ref{phase2} to process their raw data before sending them into Phase~\ref{phase3}. In Barnum~\cite{yagemann2019barnum}, the instruction sequences from program run-time tracing were sliced into basic-blocks. Then, they assigned each basic-blocks with an unique basic-block ID (BBID). Finally, due to the nature of control-flow hijacking attack, they selected the sequences ending with indirect branch instruction (e.g., indirect call/jump, return and so on) as the training data. In CNNoverCFG~\cite{phan2017convolutional}, each of instructions in CFG were labeled with its attributes in multiple perspectives, such as opcode, operands, and the function it belongs to. The training data is generated are sequences generated by traversing the attributed control-flow graph. Nguyen and others~\cite{nguyen2018auto} converted the lazy-binding CFG to corresponding adjacent matrix and treated the matrix as a image as their training data.

\item  \observation{} \textit{All the papers in our survey did not adopt Phase~\ref{phase3}.}

We observed all the papers we surveyed did not adopted Phase~\ref{phase3}. Instead, they adopted the form of numerical representation directly as their training data. Specifically, Barnum\cite{yagemann2019barnum} grouped the the instructions into basic-blocks, then represented basic-blocks with uniquely assigning IDs. In CNNoverCFG~\cite{phan2017convolutional}, each of instructions in the CFG was represented by a vector that associated with its attributes. Nguyen and others directly used the hashed value of bit string representation. 

\item  \observation{} \textit{Various Phase IV models were used.}

Barnum\cite{yagemann2019barnum} utilized BBID sequence to monitor the execution flow of the target program, which is sequence-type data. Therefore, they chose LSTM architecture to better learn the relationship between instructions. While in the other two papers\cite{phan2017convolutional,nguyen2018auto}, they trained CNN and directed graph-based CNN to extract information from control-flow graph and image, respectively.

\end{itemize}
 
The above observations seem to indicate the following indications: 
\begin{itemize}[label={}]
\item \indication{} \textit{All the existing works did not achieve Just-In-Time CFI violation detection.} 

It is still a challenge to tightly embed Deep Learning model in program execution. All existing work adopted lazy-checking~\textendash~checking the program's execution trace following its execution.

\item \indication{} \textit{There is no unified opinion on how to generate malicious sample.}

Data are hard to collect in control-flow hijacking attacks. The researchers must carefully craft malicious sample. 
It is not clear whether the ``handcrafted'' sample can reflect the nature the control-flow hijacking attack. 

\item \indication{*} The choice of methods in  \textit{Phase~\ref{phase2} are based on researchers’ security domain knowledge.}

\end{itemize}

{\color{myblack}
\subsection{Discussion}
The strength of using deep learning to solve CFI problems is that it can avoid the complicated processes of developing algorithms to build acceptable CFGs for the protected programs.
 Compared with the traditional approaches, 
 the DL based method could prevent CFI designer from 
 studying the language features of the targeted program and could also avoid the open problem (pointer analysis) in control flow analysis. 
 Therefore, DL based CFI provides us a more generalized, scalable, and secure solution. 
 However, since using DL in CFI problem is still at an early age, 
 which kinds of control-flow related data are more effective is still unclear yet in this research area. 
 Additionally, applying DL in real-time control-flow violation detection remains an untouched area and needs further research.
}

\section{A closer look at applications of Deep Learning in defending network attacks}
\label{sec:network}
\subsection{Introduction}
Network security is becoming more and more important as we depend more and more on networks for our daily lives, works and researches. Some common network attack types include probe, denial of service (DoS), Remote-to-local (R2L), etc. Traditionally, people try to detect those attacks using signatures, rules, and unsupervised anomaly detection algorithms. 
{\color{myblack}However, signature based methods can be easily fooled by slightly changing the attack payload; rule based methods need experts to regularly update rules; and unsupervised anomaly detection algorithms tend to raise lots of false positives.}
Recently, people are trying to apply Deep Learning methods for network attack detection.

In this section, we will review the very recent seven representative works that use Deep Learning for defending network attacks. \cite{millar2018deep,varenne2019intelligent,ustebay2019cyber} build neural networks for multi-class classification, whose class labels include one benign label and multiple malicious labels for different attack types. \cite{zhang2019pccn} ignores normal network activities and proposes parallel cross convolutional neural network (PCCN) to classify the type of malicious network activities. \cite{yuan2017deepdefense} applies Deep Learning to detecting a specific attack type, distributed denial of service (DDoS) attack. \cite{yin2017deep,faker2019intrusion} explores both binary classification and multi-class classification for benign and malicious activities. Among these seven works, we select two representative works~\cite{millar2018deep,zhang2019pccn} and summarize the main aspects of their approaches regarding whether the four phases exist in their works, and what exactly do they do in the Phase if it exists. We direct interested readers to Table~\ref{Table:Summary} for a concise overview of these two works.



Our review will be centered around three questions described in Section~\ref{threequestions}. In the remaining of this section, we will first provide a set of observations, and then we provide the indications. Finally, we provide some general remarks.

\subsection{Key findings from a closer look}
From a close look at the very recent applications using Deep Learning for solving network attack challenges, we observed the followings: 

\begin{itemize}[label={}]
\item  \observation{} \textit{All the seven works in our survey used public datasets, such as UNSW-NB15~\cite{moustafa2015unsw} and CICIDS2017~\cite{CICIDS2017}.} 

The public datasets were all generated in test-bed environments, with unbalanced simulated benign and attack activities. For attack activities, the dataset providers launched multiple types of attacks, and the numbers of malicious data for those attack activities were also unbalanced. 

\item  \observation{} \textit{The public datasets were given into one of two data formats, i.e., PCAP and CSV.} 

One was raw PCAP or parsed CSV format, containing network packet level features, and the other was also CSV format, containing network flow level features, which showed the statistic information of many network packets. Out of all the seven works, \cite{yuan2017deepdefense,varenne2019intelligent} used packet information as raw inputs, \cite{yin2017deep,zhang2019pccn,ustebay2019cyber,faker2019intrusion} used flow information as raw inputs, and \cite{millar2018deep} explored both cases.

\item  \observation{} \textit{In order to parse the raw inputs, preprocessing methods, including one-hot vectors for categorical texts, normalization on numeric data, and removal of unused features/data samples, were commonly used.} 

Commonly removed features include IP addresses and timestamps. \cite{faker2019intrusion} also removed port numbers from used features. By doing this, they claimed that they could ``avoid over-fitting and let the neural network learn characteristics of packets themselves''. One outlier was that, when using packet level features in one experiment, \cite{millar2018deep} blindly chose the first 50 bytes of each network packet without any feature extracting processes and fed them into neural network.

\item  \observation{} \textit{Using image representation improved the performance of security solutions using Deep Learning.}

After preprocessing the raw data, while \cite{zhang2019pccn} transformed the data into image representation, \cite{yuan2017deepdefense,varenne2019intelligent,faker2019intrusion,ustebay2019cyber,yin2017deep} directly used the original vectors as an input data. Also, \cite{millar2018deep} explored both cases and reported better performance using image representation.

\item  \observation{} \textit{None of all the seven surveyed works considered representation learning.} 

All the seven surveyed works belonged to class 1 shown in Figure~\ref{fig:dec}. They either directly used the processed vectors to feed into the neural networks, or changed the representation without explanation. One research work~\cite{millar2018deep} provided a comparison on two different representations (vectors and images) for the same type of raw input. However, the other works applied different preprocessing methods in Phase~\ref{phase2}. That is, since the different preprocessing methods generated different feature spaces, it was difficult to compare the experimental results. 

\item  \observation{} \textit{Binary classification model showed better results from most experiments.}

Among all the seven surveyed works, \cite{yuan2017deepdefense} focused on one specific attack type and only did binary classification to classify whether the network traffic was benign or malicious. Also, \cite{millar2018deep,ustebay2019cyber,zhang2019pccn,varenne2019intelligent} included more attack types and did multi-class classification to classify the type of malicious activities, and \cite{yin2017deep,faker2019intrusion} explored both cases. As for multi-class classification, the accuracy for selective classes was good, while accuracy for other classes, usually classes with much fewer data samples, suffered by up to 20\% degradation.

\item  \observation{} \textit{Data representation influenced on choosing a neural network model.} 

\end{itemize}

The above observations seem to indicate the following indications: 
\begin{itemize}[label={}]
\item \indication{} \textit{All works in our survey adopt a kind of preprocessing methods in Phase~\ref{phase2}, because raw data provided in the public datasets are either not ready for neural networks, or that the quality of data is too low to be directly used as data samples.}

Preprocessing methods can help increase the neural network performance by improving the data samples' qualities. Furthermore, by reducing the feature space, pre-processing can also improve the efficiency of neural network training and testing. Thus, Phase~\ref{phase2} should not be skipped. If Phase~\ref{phase2} is skipped, the performance of neural network is expected to go down considerably.

\item \indication{} \textit{Although Phase~\ref{phase3} is not employed in any of the seven surveyed works, none of them explains a reason for it. Also, they all do not take representation learning into consideration.}

\item \indication{} \textit{Because no work uses representation learning, the effectiveness are not well-studied.} 

Out of other factors, it seems that the choice of pre-processing methods has the largest impact, because it directly affects the data samples fed to the neural network.

\item \indication{} \textit{There is no guarantee that CNN also works well on images converted from network features.} 

Some works that use image data representation use CNN in Phase~\ref{phase4}. Although CNN has been proven to work well on image classification problem in the recent years, there is no guarantee that CNN also works well on images converted from network features.

\end{itemize}

From the observations and indications above, we hereby present two recommendations: (1) Researchers can try to generate their own datasets for the specific network attack they want to detect. As stated, the public datasets have highly unbalanced number of data for different classes. Doubtlessly, such unbalance is the nature of real world network environment, in which normal activities are the majority, but it is not good for Deep Learning. \cite{varenne2019intelligent} tries to solve this problem by oversampling the malicious data, but it is better to start with a balanced data set. (2) Representation learning should be taken into consideration. Some possible ways to apply representation learning include: (a) apply word2vec method to packet binaries, and categorical numbers and texts; (b) use K-means as one-hot vector representation instead of randomly encoding texts. We suggest that any change of data representation may be better justified by explanations or comparison experiments.
{\color{myblack}
\subsection{Discussion}
One critical challenge in this field is the lack of high-quality data set suitable for applying deep learning. 
Also, there is no agreement on how to apply domain knowledge into training deep learning models for network security problems. 
Researchers have been using different pre-processing methods, data representations and model types, 
but few of them have enough explanation on why such methods/representations/models are chosen, especially for data representation.
}

\section{A closer look at applications of Deep Learning in malware classification}
\label{sec:malware}
\subsection{Introduction}
The goal of malware classification is to identify malicious behaviors in software with static and dynamic features like control-flow graph and system API calls. 
Malware and benign programs can be collected from open datasets and online websites. 
{\color{myblack}
Both the industry and the academic communities have provided approaches to detect malware with static and dynamic analyses. 
Traditional methods such as behavior-based signatures, dynamic taint tracking, and static data flow analysis require experts to manually investigate unknown files. 
However, those hand-crafted signatures are not sufficiently effective because attackers can rewrite and reorder the malware. 
Fortunately, neural networks can automatically detect large-scale malware variants with superior classification accuracy.
}

In this section, we will review the very recent twelve representative works that use Deep Learning for malware classification \cite{DeLaRosa2018,saxe2015deep,Kolosnjaji2017,McLaughlin2017,Tobiyama2016,Dahl2013,Nix2017,Kalash2018,Cui2018,David2015,Rosenberg2018,Xu2018}. \cite{DeLaRosa2018} selects three different kinds of static features to classify malware. \cite{saxe2015deep,Kolosnjaji2017,McLaughlin2017} also use static features from the PE files to classify programs. \cite{Tobiyama2016} extracts behavioral feature images using RNN to represent the behaviors of original programs.\cite{Dahl2013} transforms malicious behaviors using representative learning without neural network. \cite{Nix2017} explores RNN model with the API calls sequences as programs' features.  \cite{Cui2018, Kalash2018} skip Phase~\ref{phase2} by directly transforming the binary file to image to classify the file. \cite{David2015, Rosenberg2018} applies dynamic features to analyze malicious features. \cite{Xu2018} combines static features and dynamic features to represent programs' features. Among these works, we select two representative works \cite{DeLaRosa2018, Rosenberg2018} and identify four phases in their works shown as Table~\ref{Table:Summary}. 

Our review will be centered around three questions described in Section~\ref{threequestions}. In the remaining of this section, we will first provide a set of observations, and then we provide the indications. Finally, we provide some general remarks.

\subsection{Key findings from a closer look}
From a close look at the very recent applications using Deep Learning for solving malware classification challenges, we observed the followings: 

\begin{itemize}[label={}]

\item \observation{} \textit{Features selected in malware classification were grouped into three categories: static features, dynamic features, and hybrid features.}

Typical static features include metadata, PE import Features, Byte/Entorpy, String, and Assembly Opcode Features derived from the PE files  \cite{Kolosnjaji2017,McLaughlin2017, saxe2015deep}. De LaRosa, Kilgallon, et al.\cite{DeLaRosa2018} took three kinds of static features: byte-level, basic-level ( strings in the file, the metadata table, and the import table of the PE header), and assembly features-level. Some works directly considered binary code as static features\cite{Cui2018, Kalash2018}.

Different from static features, dynamic features were extracted by executing the files to retrieve their behaviors during execution. The behaviors of programs, including the API function calls, their parameters, files created or deleted, websites and ports accessed, etc, were recorded by a sandbox as dynamic features\cite{David2015}. The process behaviors including operation name and their result codes were extracted \cite{Tobiyama2016}. The process memory, tri-grams of system API calls and one corresponding input parameter were chosen as dynamic features\cite{Dahl2013}. An API calls sequence for an APK file was another representation of dynamic features~\cite{Nix2017, Rosenberg2018}. 

Static features and dynamic features were combined as hybrid features~\cite{Xu2018}. For static features, Xu and others in~\cite{Xu2018} used permissions, networks, calls, and providers, etc. For dynamic features, they used system call sequences.

\item  \observation{} \textit{In most works, Phase~\ref{phase2} was inevitable because extracted features needed to be vertorized for Deep Learning models.}

One-hot encoding approach was frequently used to vectorize features\cite{Kolosnjaji2017, McLaughlin2017, Rosenberg2018, Tobiyama2016, Nix2017}. Bag-of-words (BoW) and \textit{n}-gram were also considered to represent features \cite{Nix2017}. Some works brought the concepts of word frequency in NLP to convert the sandbox file to fixed-size inputs\cite{David2015}. Hashing features into a fixed vector was used as an effective method to represent features\cite{saxe2015deep}. Bytes histogram using the bytes analysis and bytes-entropy histogram with a sliding window method were considered~\cite{DeLaRosa2018}. In \cite{DeLaRosa2018}, De La Rosa and others embeded strings by hashing the ASCII strings to a fixed-size feature vector. For assembly features, they extracted four different levels of granularity: operation level (instruction-flow-graph), block level (control-flow-graph), function level (call-graph), and global level (graphs summarized). bigram, trigram and four-gram vectors and \textit{n}-gram graph were used for the hybrid features \cite{Xu2018}. 

\item  \observation{} \textit{Most Phase~\ref{phase3} methods were classified into class 1.}

Following the classification tree shown in Figure~\ref{fig:dec}, most works were classified into class 1 shown in Figure~\ref{fig:dec} except two works\cite{Dahl2013, Tobiyama2016}, which belonged to class 3 shown in Figure~\ref{fig:dec}. To reduce the input dimension, Dahl et al.\cite{Dahl2013} performed feature selection using mutual information and random projection. Tobiyama et al. generated behavioral feature images using  RNN\cite{Tobiyama2016}.

\item {\bf Observation 4:} \textit{After extracting features, two kinds of neural network architectures, i.e., one single neural network and multiple neural networks with a combined loss function, were used.}

Hierarchical structures, like convolutional layers, fully connected layers and classification layers, were used to classify programs \cite{McLaughlin2017, Dahl2013, Nix2017, saxe2015deep, Tobiyama2016, Cui2018, Kalash2018}. A deep stack of denoising autoencoders was also introduced to learn programs' behaviors \cite{David2015}. De La Rosa and others\cite{DeLaRosa2018} trained three different models with different features to compare which static features are relevant for the classification model. Some works investigated LSTM models for sequential features~\cite{Nix2017, Rosenberg2018}. 

Two networks with different features as inputs were used for malware classification by combining their outputs with a dropout layer and an output layer\cite{Kolosnjaji2017}. In \cite{Kolosnjaji2017}, one network transformed PE Metadata and import features using feedforward neurons, another one leveraged convolutional network layers with opcode sequences. Lifan Xu et al.\cite{Xu2018} constructed a few networks and combined them using a two-level multiple kernel learning algorithm.

\end{itemize}
 
The above observations seem to indicate the following indications: 
\begin{itemize}[label={}]
\item \indication{} \textit{Except two works transform binary into images\cite{Cui2018, Kalash2018}, most works surveyed need to adapt methods to vectorize extracted features.} 

The vectorization methods should not only keep syntactic and semantic information in features, but also consider the definition of the Deep Learning model. 

\item \indication{} \textit{Only limited works have shown how to transform features using representation learning.} 

Because some works assume the dynamic and static sequences, like API calls and instruction, and have similar syntactic and semantic structure as natural language, some representation learning techniques like word2vec may be useful in malware detection. In addition, for the control-flow graph, call graph and other graph representations, graph embedding is a potential method to transform those features.

\end{itemize} 

{\color{myblack}
\subsection{Discussion}
Though several pieces of research have been done in malware detection using Deep Learning, 
it's hard to compare their methods and performances because of two uncertainties in their approaches. 
First, the Deep Learning model is a black-box, researchers cannot detail which kind of features the model learned and explain why their model works. 
Second, feature selection and representation affect the model’s performance. Because they do not use the same datasets, 
researchers cannot prove their approaches~\textendash~including selected features and Deep Learning model~\textendash~are better than others. 
The reason why few researchers use open datasets is that existing open malware datasets are out of data and limited. 
Also, researchers need to crawl benign programs from app stores, so their raw programs will be diverse.
}

\section{A closer look at applications of Deep Learning in 
system-event-based anomaly detection}
\label{sec:anomaly}
\subsection{Introduction}
System logs recorded significant events at various critical points, which can be used to debug the system's performance issues and failures. 
Moreover, log data are available in almost all computer systems and are a valuable resource for understanding system status. 
{\color{myblack}
There are a few challenges in anomaly detection based on system logs. 
Firstly, the raw log data are unstructured, while their formats and semantics can vary significantly. 
Secondly, logs are produced by concurrently running tasks. 
Such concurrency makes it hard to apply workflow-based anomaly detection methods. 
Thirdly, logs contain rich information and complexity types, including text, real value, IP address, timestamp, and so on. 
The contained information of each log is also varied. 
Finally, there are massive logs in every system. 
Moreover, each anomaly event usually incorporates a large number of logs generated in a long period.  
}

Recently, a large number of scholars employed deep learning techniques \cite{Du:2017:deeLog,loganomaly:2019,Das:2018,Brown:2018,Zhang:2019,Bertero:8109100} 
to detect anomaly events in the system logs and diagnosis system failures. 
The raw log data are unstructured, while their formats and semantics can vary significantly. 
To detect the anomaly event, the raw log usually should be parsed to structure data, 
the parsed data can be transformed into a representation that supports an effective deep learning model. 
Finally, the anomaly event can be detected by deep learning based classifier or predictor.

In this section, we will review the very recent six representative papers that use deep learning for system-event-based anomaly detection\cite{Du:2017:deeLog,loganomaly:2019,Das:2018,Brown:2018,Zhang:2019,Bertero:8109100}. DeepLog \cite{Du:2017:deeLog} utilizes LSTM to model the system log as a natural language sequence, which automatically learns log patterns from the normal event, and detects anomalies when log patterns deviate from the trained model. LogAnom\cite{loganomaly:2019} employs Word2vec to extract the semantic and syntax information from log templates. Moreover, it uses sequential and quantitative features simultaneously. Desh \cite{Das:2018} uses LSTM to predict node failures that occur in super computing systems from HPC logs. Andy Brown et al. \cite{Brown:2018} presented RNN language models augmented with attention for anomaly detection in system logs. LogRobust \cite{Zhang:2019} uses FastText to represent semantic information of log events, which can identify and handle unstable log events and sequences. Christophe Bertero et al.\cite{Bertero:8109100} map log word to a high dimensional metric space using Google's word2vec algorithm and take it as features to classify. Among these six papers, we select two representative works \cite{Du:2017:deeLog,loganomaly:2019} and summarize the four phases of their approaches. We direct interested readers to Table~\ref{Table:Summary} for a concise overview of these two works.

Our review will be centered around three questions described in Section~\ref{threequestions}. In the remaining of this section, we will first provide a set of observations, and then we provide the indications. Finally, we provide some general remarks.

\subsection{Key findings from a closer look}
From a close look at the very recent applications using deep learning for solving security-event-based anomaly detection challenges, we observed the followings: 

\begin{itemize}[label={}]
\item  \observation{} \textit{Most works of our surveyed papers evaluated their performance using public datasets.} 

By the time we surveyed this paper, only two works in \cite{Das:2018,Bertero:8109100} used their private datasets. 

\item  \observation{} \textit{Most works in this survey adopted Phase~\ref{phase2} when parsing the raw log data.} 

After reviewing the six works proposed recently, we found that five works\cite{Du:2017:deeLog,loganomaly:2019,Das:2018,Brown:2018,Zhang:2019} employed parsing technique, while only one work \cite{Bertero:8109100} did not.
\par DeepLog\cite{Du:2017:deeLog} parsed the raw log to different log type using Spell\cite{spell2016} which is based a longest common subsequence. Desh \cite{Das:2018} parsed the raw log to constant message and variable component. Loganom\cite{loganomaly:2019} parsed the raw log to different log templates using FT-Tree \cite{ftree2017} according to the frequent combinations of log words. Andy Brown et al. \cite{Brown:2018} parsed the raw log into word and character tokenization. LogRobust\cite{Zhang:2019} extracted its log event by abstracting away the parameters in the message. Christophe Bertero et al.\cite{Bertero:8109100} considered logs as regular text without parsing.

\item  \observation{}\label{log:obs:3} \textit{Most works have considered and adopted Phase~\ref{phase3}.} 

Among these six works, only DeepLog represented the parsed data using the one-hot vector without learning. Moreover, Loganom\cite{loganomaly:2019} compared their results with DeepLog. That is, DeepLog belongs to class 1 and Loganom belongs to class 4 in Figure \ref{fig:dec}, while the other four works follow in class 3.
\par The four works \cite{loganomaly:2019,Das:2018, Zhang:2019, Bertero:8109100} used word embedding techniques to represent the log data. Andy Brown et al. \cite{Brown:2018} employed attention vectors to represent the log messages. 
\par DeepLog \cite{Du:2017:deeLog} employed the one-hot vector to represent the log type without learning. We have engaged an experiment replacing the one-hot vector with trained word embeddings. 

\item  \observation{} \textit{Evaluation results were not compared using the same dataset.}

DeepLog \cite{Du:2017:deeLog} employed the one-hot vector to represent the log type without learning, which employed Phase~\ref{phase2} without Phase~\ref{phase3}. However, Christophe Bertero et al.\cite{Bertero:8109100} considered logs as regular text without parsing, and used Phase~\ref{phase3} without Phase~\ref{phase2}. The precision of the two methods is very high, which is greater than 95\%. Unfortunately, the evaluations of the two methods used different datasets.

\item  \observation{}\label{log:obs:5} \textit{Most works empolyed LSTM in Phase~\ref{phase4}.}

Five works including\cite{Du:2017:deeLog,loganomaly:2019,Das:2018,Brown:2018,Zhang:2019} employed LSTM in the Phase~\ref{phase4}, while Christophe Bertero et al.\cite{Bertero:8109100} tried different classifiers including naive Bayes, neural networks and random forest.
\end{itemize}
 
The above observations seem to indicate the following indications: 
\begin{itemize}[label={}]
\item \indication{} \textit{Phase~\ref{phase2} has a positive effect on accuracy if being well-designed.}

Since Christophe Bertero et al.~\cite{Bertero:8109100} considers logs as regular text without parsing, we can say that Phase~\ref{phase2} is not required. However, we can find that most of the scholars employed parsing techniques to extract structure information and remove the useless noise.

\item \indication{} \textit{Most of the recent works use trained representation to represent parsed data.} 

As shown in Table \ref{tab:deeplog}, we can find Phase~\ref{phase3} is very useful, which can improve detection accuracy.

\item \indication{} \textit{Phase~\ref{phase2} and Phase~\ref{phase3} cannot be skipped simultaneously.} 

Both Phase~\ref{phase2} and Phase~\ref{phase3} are not required. However, all methods have employed Phase~\ref{phase2} or Phase~\ref{phase3}.

\item \indication{} \textit{Observation~8.\ref{log:obs:3} indicates that the trained word embedding format can improve the anomaly detection accuracy as shown in Table~\ref{tab:deeplog}.}

\begin{center}
\normalsize
  \begin{threeparttable}
  \caption{Comparison between word embedding and one-hot representation.}
  \label{tab:deeplog}
\begin{tabular}{p{3.5cm}||p{1.5cm}p{1.5cm}p{2cm}p{2cm}p{2cm}}  
\hline  
\hline  
Method  &  FP~\tnote{1}  & FN~\tnote{2}  & Precision & Recall& F1-measure\\ 
\hline  
Word Embedding~\tnote{3} & 680 & 219 &   96.069\%  & 98.699\% & 97.366\% \\ 
\hline  
One-hot Vector~\tnote{4} & 711 & 705 &   95.779\%  & 95.813\% & 95.796\% \\ 
\hline  
DeepLog~\tnote{5} & 833 & 619 &   95\%  & 96\% & 96\% \\ 
\hline 
\hline   
\end{tabular} 
    \begin{tablenotes}
        \item \tnote{1}FP: false positive; \tnote{2}FN: False negative;\tnote{3}Word Embedding: Log keys are embedded by Continuous Bag of words;\tnote{4} One-hot Vector: We reproduced the results according to DeepLog;\tnote{5} DeepLog: Orignial results presented in the paper \cite{Du:2017:deeLog}.
    \end{tablenotes}
\end{threeparttable}
\end{center}

\item \indication{} \textit{Observation~8.\ref{log:obs:5} indicates that most of the works adopt LSTM to detect anomaly events.}

We can find that most of the works adopt LSTM to detect anomaly event, since log data can be considered as sequence and there can be lags of unknown duration between important events in a time series. LSTM has feedback connections, which can not only process single data points, but also entire sequences of data.

\end{itemize}

As our consideration, neither Phase~\ref{phase2} nor Phase~\ref{phase3} is required in system event-based anomaly detection. However, Phase~\ref{phase2} can remove noise in raw data, and Phase~\ref{phase3} can learn a proper representation of the data. Both Phase~\ref{phase2} and Phase~\ref{phase3} have a positive effect on anomaly detection accuracy. Since the event log is text data that we can't feed the raw log data into deep learning model directly, Phase~\ref{phase2} and Phase~\ref{phase3} can't be skipped simultaneously.

{\color{myblack}
\subsection{Discussion}
Deep learning can capture the potentially nonlinear and high dimensional dependencies among log entries from the training data that correspond to abnormal events. 
In that way, it can release the challenges mentioned above. 
However, it still suffers from several challenges. 
For example, how to represent the unstructured data accurately and automatically without human knowledge. 
}

\section{A closer look at applications of Deep Learning in solving memory forensics challenges}
\label{sec:forensic}
\subsection{Introduction}
In the field of computer security, memory forensics is security-oriented forensic analysis of a computer’s memory dump. 
Memory forensics can be conducted against OS kernels, user-level applications, as well as mobile devices. 
Memory forensics outperforms traditional disk-based forensics because although secrecy attacks can erase their footprints on disk, 
they would have to appear in memory \cite{Song:2018:DeepMem}. 
The memory dump can be considered as a sequence of bytes, 
thus memory forensics usually needs to extract security semantic information from raw memory dump to find attack traces. 

{\color{myblack} 
The traditional memory forensic tools fall into two categories: signature scanning and data structure traversal. 
These traditional methods usually have some limitations. 
Firstly, it needs expert knowledge on the related data structures to create signatures or traversing rules. 
Secondly, attackers may directly manipulate data and pointer values in kernel objects to evade detection, 
and then it becomes even more challenging to create signatures and traversing rules that cannot be easily violated by malicious manipulations, system updates, and random noise. 
Finally, the high-efficiency requirement often sacrifices high robustness. 
For example, an efficient signature scan tool usually skips large memory regions that are unlikely to have the relevant objects and relies on simple but easily tamperable string constants. 
An important clue may hide in this ignored region.
} 

In this section, we will review the very recent four representative works that use Deep Learning for memory forensics\cite{Song:2018:DeepMem,Petrik:2018,Michalas:2017,DAI201830}. DeepMem\cite{Song:2018:DeepMem} recognized the kernel objects from raw memory dumps by generating abstract representations of kernel objects with a graph-based Deep Learning approach. MDMF\cite{Petrik:2018} detected OS and architecture-independent malware from memory snapshots with several pre-processing techniques, domain unaware feature selection, and a suite of machine learning algorithms. MemTri\cite{Michalas:2017} predicts the likelihood of criminal activity in a memory image using a Bayesian network, based on evidence data artefacts generated by several applications. Dai et al. \cite{DAI201830} monitor the malware process memory and classify malware according to memory dumps, by transforming the memory dump into grayscale images and adopting a multi-layer perception as the classifier. 

Among these four works\cite{Song:2018:DeepMem,Petrik:2018,Michalas:2017,DAI201830}, two representative works (i.e., \cite{Song:2018:DeepMem,Petrik:2018}) are already summarized phase-by-phase in Table 1. We direct interested readers to Table~\ref{Table:Summary} for a concise overview of these two works.

Our review will be centered around the three questions raised in Section~\ref{threequestions}. In the remaining of this section, we will first provide a set of observations, and then we provide the indications. Finally, we provide some general remarks.

\subsection{Key findings from a closer look}
From a close look at the very recent applications using Deep Learning for solving memory forensics challenges, we observed the followings: 

\begin{itemize}[label={}]
\item  \observation{}\label{mem:obs1} \textit{Most methods used their own datasets for performance evaluation, while none of them used a public dataset.} 

DeepMem was evaluated on self-generated dataset by the authors, who collected a large number of diverse memory dumps, and labeled the kernel objects in them using existing memory forensics tools like Volatility. MDMF employed the MalRec dataset by Georgia Tech to generate malicious snapshots, while it created a dataset of benign memory snapshots running normal software.
MemTri ran several Windows 7 virtual machine instances with self-designed suspect activity scenarios to gather memory images. Dai et al. built the Procdump program in Cuckoo sandbox to extract malware memory dumps. 
We found that each of the four works in our survey generated their own datasets, while none was evaluated on a public dataset. 

\item  \observation{}\label{mem:obs2} \textit{Among the four works\cite{Song:2018:DeepMem,Michalas:2017,Petrik:2018,DAI201830}, two works \cite{Song:2018:DeepMem,Michalas:2017} employed Phase~\ref{phase2} while the other two works \cite{Petrik:2018,DAI201830} did not employ.}

DeepMem \cite{Song:2018:DeepMem} devised a graph representation for a sequence of bytes, taking into account both adjacency and points-to relations, to better model the contextual information in memory dumps. MemTri \cite{Michalas:2017} firstly identified the running processes within the memory image that match the target applications, then employed regular expressions to locate evidence artefacts in a memory image. MDMF \cite{Petrik:2018} and Dai et al. \cite{DAI201830} transformed the memory dump into image directly.

\item  \observation{}\label{mem:obs3} \textit{Among four works\cite{Song:2018:DeepMem,Michalas:2017,Petrik:2018,DAI201830}, only DeepMem \cite{Song:2018:DeepMem} employed Phase~\ref{phase3} for which it used an embedding method to represent a memory graph.} 

MDMF\cite{Petrik:2018} directly fed the generated memory images into the training of a CNN model. Dai et al. \cite{DAI201830} used HOG feature descriptor for detecting objects, while MemTri\cite{Michalas:2017} extracted evidence artefacts as the input of Bayesian Network. In summary, DeepMem belonged to class 3 shown in Figure~\ref{fig:dec}, while the other three works belonged to class 1 shown in Figure~\ref{fig:dec}.

\item  \observation{}\label{mem:obs4} \textit{All the four works\cite{Song:2018:DeepMem,Petrik:2018,Michalas:2017,DAI201830} have employed different classifiers even when the types of input data are the same.}  

DeepMem chose fully connected network (FCN) model that has multi-layered hidden neurons with ReLU activation functions, following by a softmax layer as the last layer. MDMF\cite{Petrik:2018} evaluated their performance both on traditional machine learning algorithms and Deep Learning approach including CNN and LSTM. Their results showed the accuracy of different classifiers did not have a significant difference. MemTri employed a Bayesian network model that is designed with three layers, i.e., a hypothesis layer, a sub-hypothesis layer, and an evidence layer. Dai et al. used a multi-layer perception model including an input layer, a hidden layer and an output layer as the classifier. 
\end{itemize}
 
The above observations seem to indicate the following indications:
\begin{itemize}[label={}]
\item \indication{} \textit{There lacks public datasets for evaluating the performance of different Deep Learning methods in memory forensics.}

From Observation~9.\ref{mem:obs1},  we find that none of the four works surveyed was evaluated on public datasets. 

\item \indication{} \textit{From Observation~9.\ref{mem:obs2}, we find that it is disputable whether one should employ Phase~\ref{phase2} when solving memory forensics problems.}

Since both \cite{Petrik:2018} and \cite{DAI201830} 
 directly transformed a memory dump into an image, Phase~\ref{phase2} is not required in these two works. However, since there is a large amount of useless information in a memory dump, we argue that appropriate prepossessing could improve the accuracy of the trained models. 

\item \indication{} \textit{From Observation~9.\ref{mem:obs3},  we find that Phase~\ref{phase3} is paid not much attention in memory forensics.} 

Most works did not employ Phase~\ref{phase3}. Among the four works, only DeepMem \cite{Song:2018:DeepMem} employed Phase~\ref{phase3} during which it used embeddings to represent a memory graph. The other three works  \cite{Petrik:2018,Michalas:2017,DAI201830} did not learn any representations before training a Deep Learning model.  

\item \indication{} \textit{For Phase~\ref{phase4} in memory forensics, different classifiers can be employed.}

Which kind of classifier to use seems to be determined by the features used and their data structures. From Observation~9.\ref{mem:obs4}, we find that the four works have actually employed different kinds of classifiers even the types of input data are the same. It is very interesting that MDMF obtained similar results with different classifiers including traditional machine learning and Deep Learning models. However, the other three works did not discuss why they chose a particular kind of classifier. 
\end{itemize}

Since a memory dump can be considered as a sequence of bytes, the data structure of a training data example is straightforward. If the memory dump is transformed into a simple form in Phase~\ref{phase2}, it can be directly fed into the training process of a Deep Learning model, and as a result Phase~\ref{phase3} can be ignored. However, if the memory dump is transformed into a complicated form in Phase~\ref{phase2}, Phase~\ref{phase3} could be quite useful in memory forensics. 

Regarding the answer for Question 3 at Section~\ref{threequestions}, it is very interesting that during Phase~\ref{phase4} different classifiers can be employed in memory forensics. Moreover, MDMF \cite{Petrik:2018} has shown that they can obtain similar results with different kinds of classifiers. Nevertheless, they also admit that with a larger amount of training data, the performance could be improved by Deep Learning. 

{\color{myblack}
\subsection{Discussion}
An end-to-end manner deep learning model can learn the precise representation of memory dump automatically to release the requirement for expert knowledge. 
However, it still needs expert knowledge to represent data and attacker behavior. 
Attackers may also directly manipulate data and pointer values in kernel objects to evade detection.
}

\section{A closer look at applications of Deep Learning in security-oriented fuzzing}
\label{sec:fuzzing}
\subsection{Introduction}
Fuzzing of software security is one of the state of art techniques that people use to detect software vulnerabilities. The goal of fuzzing is to find all the vulnerabilities exist in the program by testing as much program code as possible. Due to the nature of fuzzing, this technique works best on finding vulnerabilities in programs that take in input files, like PDF viewers\cite{godefroid2017learn} or web browsers. 
A typical workflow of fuzzing can be concluded as: given several seed input files, the fuzzer will mutate or fuzz the seed inputs to get more input files, with the aim of expanding the overall code coverage of the target program as it executes the mutated files. 
{\color{myblack} Although there have already been various popular fuzzers\cite{li2018fuzzing}, fuzzing still cannot bypass its problem of sometimes redundantly testing input files which cannot improve the code coverage rate\cite{shi2019neuzz,rajpal2017not}. 
Some input files mutated by the fuzzer even cannot pass the well-formed file structure test\cite{godefroid2017learn}.} Recent research has come up with ideas of applying Deep Learning in the process of fuzzing to solve these problems.

In this section, we will review the very recent four representative works that use Deep Learning for fuzzing for software security.  Among the three, two representative works\cite{godefroid2017learn,shi2019neuzz} are already summarized phase-by-phase in Table \ref{Table:Summary}. We direct interested readers to Table \ref{Table:Summary} for a concise overview of those two works. 

Our review will be centered around three questions described in Section~\ref{threequestions}. In the remaining of this section, we will first provide a set of observations, and then we provide the indications. Finally, we provide some general remarks.

\subsection{Key findings from a closer look}
From a close look at the very recent applications using Deep Learning for solving security-oriented program analysis challenges, we observed the followings: 

\begin{itemize}[label={}]
    \item  \observation{} \textit{Deep Learning has only been applied in mutation-based fuzzing.}
    
    Even though various of different fuzzing techniques, 
    including symbolic execution based fuzzing~\cite{stephens2016driller}, tainted analysis based fuzzing~\cite{6200194} and hybrid fuzzing~\cite{217563} have been proposed so far, we observed that all the works we surveyed employed Deep Learning method to assist the primitive fuzzing~\textendash~mutation-based fuzzing.
    Specifically, they adopted Deep Learning to assist fuzzing tool's input mutation. We found that they commonly did it in two ways: 1) training Deep Learning models to tell how to efficiently mutate the input to trigger more execution path~\cite{shi2019neuzz, rajpal2017not};
    2) training Deep Learning models to tell how to keep the mutated files compliant with the program's basic semantic requirement~\cite{godefroid2017learn}.
    Besides, all three works trained different Deep Learning models for different programs,
    which means that knowledge learned from one programs cannot be applied to other programs.
    
    \item  \observation{} \textit{Similarity among all the works in our survey existed when choosing the training samples in Phase I.}
    
    The works in this survey had a common practice, i.e., using the input files directly as training samples of the Deep Learning model. Learn\&Fuzz\cite{godefroid2017learn} used character-level PDF objects sequence as training samples. Neuzz\cite{shi2019neuzz} regarded input files directly as byte sequences and fed them into the neural network model. Mohit Rajpal et al.\cite{rajpal2017not} also used byte level representations of input files as training samples.

    \item  \observation{} \textit{Difference between all the works in our survey existed when assigning the training labels in Phase I.}
    
    Despite the similarity of training samples researchers decide to use, there was a huge difference in the training labels that each work chose to use. 
    Learn\&Fuzz\cite{godefroid2017learn} directly used the character sequences of PDF objects as labels, same as training samples, but shifted by one position, which is a common generative model technique already broadly used in speech and handwriting recognition. 
    Unlike Learn\&Fuzz, Neuzz\cite{shi2019neuzz} and Rajpal’s work\cite{rajpal2017not} used bitmap and heatmap respectively as training labels, 
    with the bitmap demonstrating the code coverage status of a certain input, 
    and the heatmap demonstrating the efficacy of flipping one or more bytes of the input file. 
    Whereas, as a common terminology well-known among fuzzing researchers, bitmap was gathered directly from the results of AFL. Heatmap used by Rajpal et al. was generated by comparing the code coverage supported by the bitmap of one seed file 
    and the code coverage supported by bitmaps of the mutated seed files. It was noted that if there is acceptable level of code coverage expansion when executing the mutated seed files, demonstrated by more ``1''s, instead of ``0''s in the corresponding bitmaps, the byte level differences among the original seed file and the mutated seed files will be highlighted. Since those bytes should be the focus of later on mutation, heatmap was used to denote the location of those bytes.
    
    Different labels usage in each work was actually due to the different kinds of knowledge each work wants to learn. For a better understanding, let us note that we can simply regard a Deep Learning model as a simulation of a ``function''. Learn\&Fuzz~\cite{godefroid2017learn} wanted to learn valid mutation of a PDF file that was compliant with the syntax and semantic requirements of PDF objects. Their model could be seen as a simulation of $f(x, \theta)=y$, where $x$ denotes sequence of characters in PDF objects and $y$ represents a sequence that are obtained by shifting the input sequences by one position. They generated new PDF object character sequences given a starting prefix once the model was trained. In Neuzz\cite{shi2019neuzz}, an NN(Neural Network) model was used to do program smoothing, which simultated a smooth surrogate function that approximated the discrete branching behaviors of the target program. $f(x, \theta)=y$, where $x$ denoted program's byte level input and $y$ represented the corresponding edge coverage bitmap. In this way, the gradient of the surrogate function was easily computed, due to NN's support of efficient computation of gradients and higher order derivatives. Gradients could then be used to guide the direction of mutation, in order to get greater code coverage. In Rajpal and others' work\cite{rajpal2017not}, they designed a model to predict good (and bad) locations to mutate in input files based on the past mutations and corresponding code coverage information. Here, the $x$ variable also denoted program's byte level input, but the $y$ variable represented the corresponding heatmap. 
    
    \item  \observation{} \textit{Various lengths of input files were handled in Phase~\ref{phase2}.}
    
    Deep Learning models typically accepted fixed length input, whereas the input files for fuzzers often held different lengths. Two different approaches were used among the three works we surveyed: splitting and padding. Learn\&Fuzz~\cite{godefroid2017learn} dealt with this mismatch by concatenating all the PDF objects character sequences together, and then splited the large character sequence into multiple training samples with a fixed size. Neuzz~\cite{shi2019neuzz} solved this problem by setting a maximize input file threshold and then, padding the smaller-sized input files with null bytes. From  additional experiments, they also found that a modest threshold gived them the best result, and enlarging the input file size did not grant them additional accuracy. Aside from preprocessing training samples, Neuzz also preprocessed training labels and reduced labels dimension by merging the edges that always appeared together into one edge, in order to prevent the multicollinearity problem, that could prevent the model from converging to a small loss value. Rajpal and others\cite{rajpal2017not} used the similar splitting mechanism as Learn\&Fuzz to split their input files into either 64-bit or 128-bit chunks. Their chunk size was determined empirically and was considered as a trainable parameter for their Deep Learning model, and their approach did not require sequence concatenating at the beginning. 
      
    \item  \observation{} \textit{All the works in our survey skipped Phase~\ref{phase3}.}
    
    According to our definition of Phase~\ref{phase3}, all the works in our survey did not consider representation learning. Therefore, all the three works\cite{godefroid2017learn,shi2019neuzz,rajpal2017not} fell into class 1 shown in Figure~\ref{fig:dec}.While as in Rajpal and others' work, they considered the numerical representation of byte sequences. They claimed that since one byte binary data did not always represent the magnitude but also state, representing one byte in values ranging from 0 to 255 could be suboptimal. They used lower level 8-bit representation.

\end{itemize}
 
The above observations seem to indicate the following indications: 
\begin{itemize}[label={}]
\item \indication{} \textit{No alteration to the input files seems to be a correct approach.}

As far as we concerned, it is due to the nature of fuzzing. That is, since every bit of the input files matters, any slight alteration to the input files could either lose important information or add redundant information for the neural network model to learn. 

\item \indication{} \textit{Evaluation criteria should be chosen carefully when judging mutation.}

Input files are always used as training samples regarding using Deep Learning technique in fuzzing problems. Through this similar action, researchers have a common desire to let the neural network mode learn how the mutated input files should look like. But the criterion of judging a input file actually has two levels: on the one hand, a good input file should be correct in syntax and semantics; on the other hand, a good input file should be the product of a useful mutation, which triggers the program to behave differently from previous execution path. This idea of a fuzzer that can generate semantically correct input file could still be a bad fuzzer at triggering new execution path was first brought up in Learn\&Fuzz\cite{godefroid2017learn}. We could see later on works trying to solve this problem by using either different training labels\cite{rajpal2017not} or use neural network to do program smoothing\cite{shi2019neuzz}. We encouraged fuzzing researchers, when using Deep Learning techniques, to keep this problem in mind, in order to get better fuzzing results. 

\item \indication{} \textit{Works in our survey only focus on \textit{local knowledge}.}

In brief, some of the existing works~\cite{shi2019neuzz,rajpal2017not} 
leveraged the Deep Learning model to learn the relation between program's input and its behavior and used the knowledge that learned from history to guide future mutation.
For better demonstration, we defined the knowledge that only applied in one program as \textit{local knowledge}.
In other words, this indicates that the \textit{local knowledge} cannot direct fuzzing on other programs.

\end{itemize} 
{\color{myblack}
\subsection{Discussion}
Corresponding to the problems conventional fuzzing has, the advantages of applying DL in fuzzing are that DL’s learning ability can ensure mutated input files follow the designated grammar rules better. 
The ways in which input files are generated are more directed, and will, therefore, guarantee the fuzzer to increase its code coverage by each mutation. 
However, even if the advantages can be clearly demonstrated by the two papers we discuss above, 
some challenges still exist, including mutation judgment challenges that are faced both by traditional fuzzing techniques and fuzzing with DL, 
and the scalability of fuzzing approaches. 

We would like to raise several interesting questions for the future researchers: 1) Can the knowledge learned from the fuzzing history of one program be applied to direct testing on other programs? 2) If the answer to question one is positive, we can suppose that \textit{global knowledge} across different programs exists? 
Then, can we train a model to extract the \textit{global knowledge}? 3) Whether it is possible to combine \textit{global knowledge} and \textit{local knowledge} when fuzzing programs?
}

\section{Discussion}
\label{sec:dis}
Using high-quality data in Deep Learning is important as much as using well-structured deep neural network architectures. That is, obtaining quality data must be an important step, which should not be skipped, even in resolving security problems using Deep Learning. So far, this study demonstrated how the recent security papers using Deep Learning have adopted data conversion (Phase~\ref{phase2}) and data representation (Phase~\ref{phase3}) on different security problems. Our observations and indications showed a clear understanding of how security experts generate quality data when using Deep Learning.

Since we did not review all the existing security papers using Deep Learning, the generality of observations and indications is somewhat limited. Note that our selected papers for review have been published recently at one of prestigious security and reliability conferences such as USENIX SECURITY, ACM CCS and so on~\cite{shin2015recognizing}-\cite{Das:2018},~\cite{Brown:2018,Zhang:2019},~\cite{Song:2018:DeepMem,Petrik:2018},~\cite{wang2019neufuzz}-\cite{rajpal2017not}. Thus, our observations and indications help to understand how most security experts have used Deep Learning to solve the well-known eight security problems from program analysis to fuzzing.

Our observations show that we should transfer raw data to synthetic formats of data ready for resolving security problems using Deep Learning through data cleaning and data augmentation and so on. Specifically, we observe that Phases~\ref{phase2} and~\ref{phase3} methods have mainly been used for the following purposes: 
\begin{itemize}
\item To clean the raw data to make the neural network (NN) models easier to interpret
\item To reduce the dimensionality of data (e.g., principle component analysis (PCA), t-distributed stochastic neighbor embedding (t-SNE))
\item To scale input data (e.g., normalization)
\item To make NN models understand more complex relationships depending on security problems (e.g. memory graphs)
\item To simply change various raw data formats into a vector format for NN models (e.g. one-hot encoding and word2vec embedding) 

\end{itemize} 


In this following, we do further discuss the question, ``What if Phase~\ref{phase2} is skipped?", rather than the question, ``Is Phase~\ref{phase3} always necessary?". 
This is because most of the selected papers do not consider Phase~\ref{phase3} methods (76\%), or adopt with no concrete reasoning (19\%). 
Specifically, we demonstrate how Phase~\ref{phase2} has been adopted according to eight security problems, different types of data, various models of NN and various outputs of NN models, 
in depth. 
Our key findings are summarized as follows:
\begin{itemize}
    \item How to fit security domain knowledge into raw data has not been well-studied yet.
    \item While raw text data are commonly parsed after embedding, raw binary data are converted using various Phase~\ref{phase2} methods.
    \item Raw data are commonly converted into a vector format to fit well to a specific NN model using various Phase~\ref{phase2} methods.
    \item Various Phase~\ref{phase2} methods are used according to the relationship between output of security problem and output of NN models.
\end{itemize}

\subsection{What if Phase~\ref{phase2} is skipped?}

From the analysis results of our selected papers for review, we roughly classify Phase~\ref{phase2} methods into the following four categories. 
\begin{itemize}
\item \textit{Embedding:} The data conversion methods that intend to convert high-dimensional discrete variables into low-dimensional continuous vectors~\cite{googledev:mlcourse}.  

\item \textit{Parsing combined with embedding: } The data conversion methods that constitute an input data into syntactic components in order to test conformability after embedding. 

\item \textit{One-hot encoding:} A simple embedding where each data belonging to a specific category is mapped to a vector of $0$s and a single $1$. Here, the low-dimension transformed vector is not managed.

\item \textit{Domain-specific data structures:} A set of data conversion methods which generate data structures capturing domain-specific knowledge for different security problems, e.g., memory graphs~\cite{Song:2018:DeepMem}. 
\end{itemize} 

\subsubsection{Findings on eight security problems}
\begin{figure}
\centering
\begin{tikzpicture}
    [
        pie chart,
        slice type={embed}{myblue}{vertical lines},
        slice type={pars}{myyellow}{grid},
        slice type={hot}{mypurple}{north west lines},
        slice type={none}{mygreen}{crosshatch},
        slice type={other}{myred}{dots},
        pie values/.style={font={\small}},
        scale=1.3
    ]
    
        \pie[yshift=2.4cm]
            {PA}{100/embed}
        \pie[xshift=2.4cm,yshift=2.4cm]%
            {ROP}{66.7/embed,33.4/hot}
        \pie[xshift=4.8cm,yshift=2.4cm]%
            {CFI}{66.7/embed,33.4/other}
        \pie[xshift=7.2cm,yshift=2.4cm]%
            {NA}{100/pars}
        \pie[]
            {MC}{41.7/embed,41.7/hot,16.7/none}
        \pie[xshift=2.4cm]%
            {SEAD}{83.4/pars,16.7/none}
        \pie[xshift=4.8cm]%
            {MF}{100/other}
        \pie[xshift=7.2cm]%
            {FUZZING}{33.4/embed,66.7/none}
    
        \legend[shift={(0cm,-1cm)}]{{Embedding}/embed, {Parsing \& Embedding}/pars}
        \legend[shift={(3cm,-1cm)}]{{One-hot}/hot, {None}/none}
        \legend[shift={(6cm,-1cm)}]{{Other}/other}
    
\end{tikzpicture}
\caption{Statistics of Phase~\ref{phase2} methods for eight security problems}
\label{fig:data2}
\end{figure}
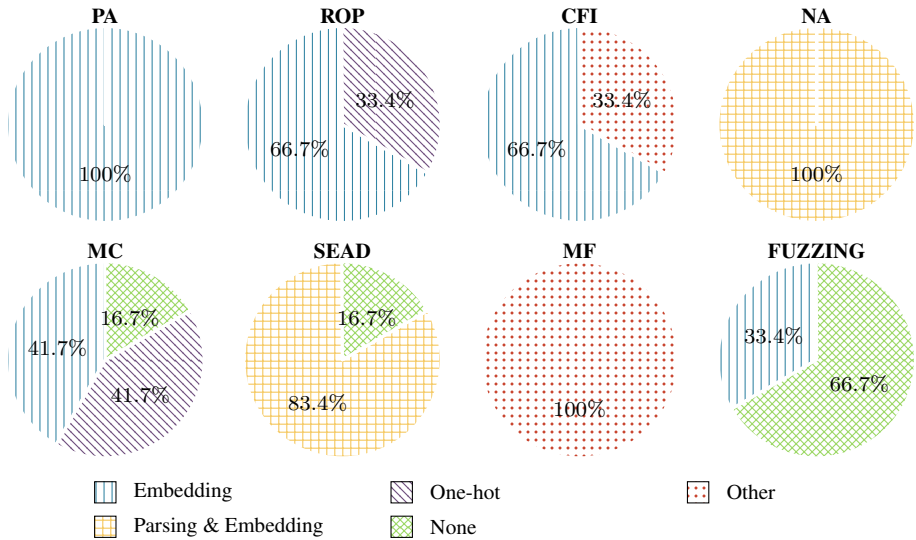

We observe that over 93\% of the papers use one of the above-classified Phase~\ref{phase2} methods. 7\% of the papers do not use any of the 
above-classified methods, and these papers are mostly solving a software fuzzing problem. 
Specifically, we observe that 35\% of the papers use a Category 1 (i.e. embedding) method; 
30\% of the papers use a Category 2 (i.e. parsing combined with embedding) method;
15\% of the papers use a Category 3 (i.e. one-hot encoding) method; 
and 13\% of the papers use a Category 4 (i.e. domain-specific data structures) method.  Regarding why one-hot encoding is not widely used, we found that 
most security data include categorical input values, which are not 
directly analyzed by Deep Learning models. 


From Figure ~\ref{fig:data2}, we also observe that according to security problems, different Phase~\ref{phase2} methods are used. First, PA, ROP and CFI should convert raw data into a vector format using embedding because they commonly collect instruction sequence from binary data. Second, NA and SEAD use parsing combined with embedding because raw data such as the network traffic and system logs consist of the complex attributes with the different formats such as categorical and numerical input values. Third, we observe that MF uses various data structures because memory dumps from memory layout are unstructured. Fourth, fuzzing generally uses no data conversion since Deep Learning models are used to generate the new input data with the same data format as the original raw data. Finally, we observe that MC commonly uses one-hot encoding and embedding because malware binary and well-structured security log files include categorical, numerical and unstructured data in general. These observations indicate that type of data strongly influences on use of Phase~\ref{phase2} methods. We also observe that only MF among eight security problems commonly transform raw data into well-structured data embedding a specialized security domain knowledge. This observation indicates that various conversion methods of raw data into well-structure data which embed various security domain knowledge are not yet studied in depth.

\subsubsection{Findings on different data types}

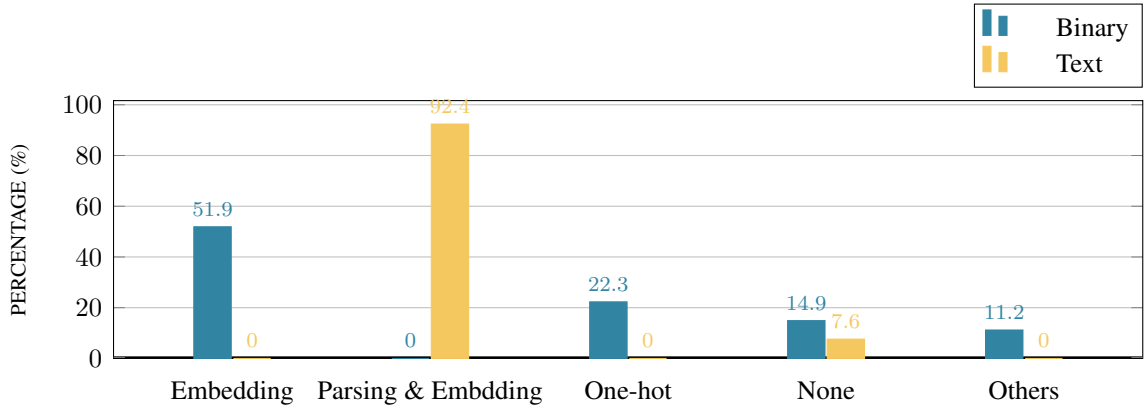
\begin{figure}[t!]
    \centering
    \begin{tikzpicture}
        \begin{axis}[
            width  = 0.95*\textwidth,
            height = 5cm,
            major x tick style = transparent,
            ybar=2*\pgflinewidth,
            bar width=14pt,
            ymajorgrids = true,
            ylabel = {PERCENTAGE (\%)},
            symbolic x coords={Embedding,Parsing \& Embdding,One-hot, None, Others},
            xtick = data,
            scaled y ticks = false,
            enlarge x limits=0.15,
            ymin=0,
            legend cell align=left,
            legend style={
                    at={(1,1.05)},
                    anchor=south east,
                    column sep=3ex,
                    font=\footnotesize,
            },
            extra y ticks = 0.4,
            extra y tick labels={},
            extra y tick style={grid=major, major grid style={thick,draw=black}},
            label style={font=\scriptsize},
            tick label style={font=\footnotesize},
            nodes near coords,
            every node near coord/.append style={font=\scriptsize},
        ]
            \addplot[style={myblue,fill=myblue,mark=none}]
                coordinates {(Embedding, 51.9) (Parsing \& Embdding,0.0) (One-hot,22.3)(None,14.9)(Others,11.2)};
    
            \addplot[style={myyellow,fill=myyellow,mark=none}]
                coordinates {(Embedding, 0.0) (Parsing \& Embdding,92.4) (One-hot,0.0)(None,7.6)(Others,0.0)};
    
            \legend{Binary,Text}
        \end{axis}
    \end{tikzpicture}
    \caption{Statistics of Phase~\ref{phase2} methods on type of data.}
    \label{fig:data1-2}
\end{figure}

Note that according to types of data, a NN model works better than the others. For example, CNN works well with images but does not work with text. From Figure ~\ref{fig:data1-2} for raw binary data, we observe that 51.9\%, 22.3\% and 11.2\% of security papers use embedding, one-hot encoding and $Others$, respectively. Only 14.9\% of security papers, especially related to fuzzing, do not use one of Phase~\ref{phase2} methods. This observation indicates that binary input data which have various binary formats should be converted into an input data type which works well with a specific NN model. From Figure ~\ref{fig:data1-2} for raw text data, we also observe that 92.4\% of papers use parsing with embedding as the Phase~\ref{phase2} method. Note that compared with raw binary data whose formats are unstructured, raw text data generally have the well-structured format. Raw text data collected from network traffics may also have various types of attribute values. Thus, raw text data are commonly parsed after embedding to reduce redundancy and dimensionality of data.

\subsubsection{Findings on various models of NN}

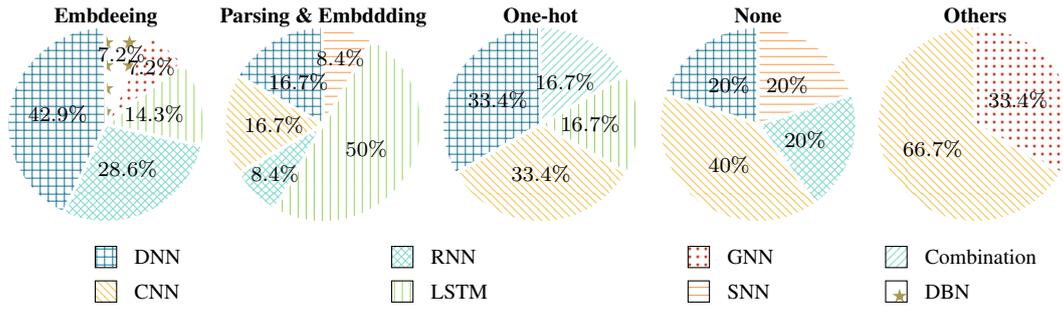
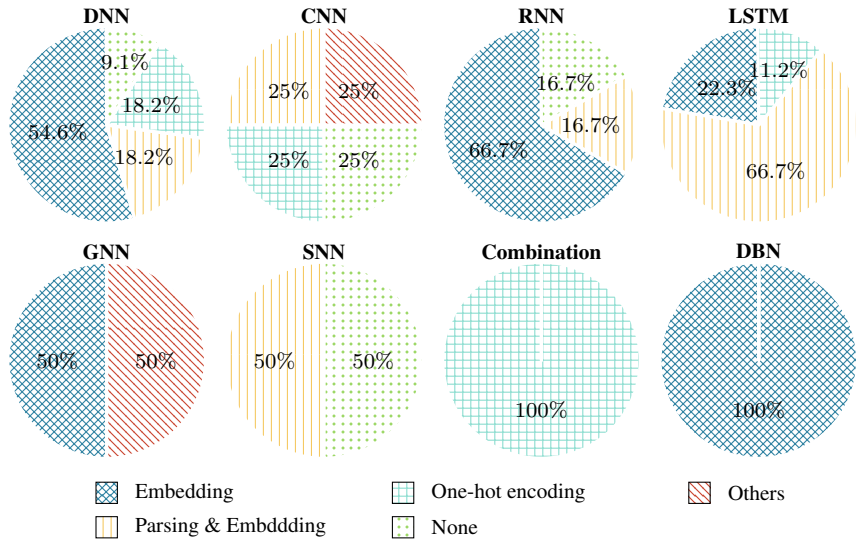
\begin{figure}[t!]
    \centering
    \begin{subfigure}[b]{0.90\textwidth}
    \begin{tikzpicture}
        [
            pie chart,
            slice type={dnn}{myblue}{grid},
            slice type={cnn}{myyellow}{north west lines},
            slice type={rnn}{mylightblue}{crosshatch},
            slice type={lstm}{mygreen}{vertical lines},
            slice type={gnn}{myred}{dots},
            slice type={snn}{myorange}{horizontal lines},
            slice type={combination}{mylightblue}{north east lines},
            slice type={dbn}{mybrown}{fivepointed stars},
            pie values/.style={font={\small}},
            scale=1.3
        ]
            \pie{Embdeeing}{42.9/dnn,28.6/rnn,14.3/lstm,7.2/gnn,7.2/dbn}
            \pie[xshift=2.2cm]%
                {Parsing \& Embddding}{16.7/dnn,16.7/cnn,8.4/rnn,50/lstm,8.4/snn}
            \pie[xshift=4.4cm]%
                {One-hot}{33.4/dnn,33.4/cnn,16.7/lstm,16.7/combination}
            \pie[xshift=6.6cm]%
                {None}{20/dnn,40/cnn,20/rnn,20/snn}
            \pie[xshift=8.8cm]
                {Others}{66.7/cnn,33.4/gnn}
            
            \legend[shift={(0cm,-1cm)}]{{DNN}/dnn, {CNN}/cnn}
            \legend[shift={(3cm,-1cm)}]{{RNN}/rnn, {LSTM}/lstm}
            \legend[shift={(6cm,-1cm)}]{{GNN}/gnn, {SNN}/snn}
            \legend[shift={(8cm,-1cm)}]{{Combination}/combination,{DBN}/dbn}
        
    \end{tikzpicture}
    \caption{Statistics of Phase~\ref{phase2} methods for eight security problems.}
    \label{fig:data2}
\end{subfigure}
\hfill
    \begin{subfigure}[b]{1.00\textwidth}
        \centering
    
        \begin{tikzpicture}
            [
                pie chart,
                slice type={embed}{myblue}{crosshatch},
                slice type={parsing}{myyellow}{vertical lines},
                slice type={onehot}{mylightblue}{grid},
                slice type={none}{mygreen}{dots},
                slice type={others}{myred}{north west lines},
                pie values/.style={font={\small}},
                scale=1.3
            ]
    
                \pie[yshift=2.4cm]
                    {DNN}{54.6/embed,18.2/parsing,18.2/onehot,9.1/none}
                \pie[xshift=2.2cm,yshift=2.4cm]%
                    {CNN}{25/parsing,25/onehot,25/none,25/others}
                \pie[xshift=4.4cm,yshift=2.4cm]%
                    {RNN}{66.7/embed,16.7/parsing,16.7/none}
                \pie[xshift=6.6cm,yshift=2.4cm]%
                    {LSTM}{22.3/embed,66.7/parsing,11.2/onehot}

                \pie{GNN}{50/embed,50/others}
                \pie[xshift=2.2cm]%
                    {SNN}{50/parsing,50/none}
                \pie[xshift=4.4cm]%
                    {Combination}{100/onehot}
                \pie[xshift=6.6cm]%
                    {DBN}{100/embed}

                \legend[shift={(0cm,-1cm)}]{{Embedding}/embed, {Parsing \& Embddding}/parsing}
                \legend[shift={(3cm,-1cm)}]{{One-hot encoding}/onehot, {None}/none}
                \legend[shift={(6cm,-1cm)}]{{Others}/others}
            
        \end{tikzpicture}
        \caption{Phase~\ref{phase2} methods over type of NN.}
        \label{fig:data1-4}
    \end{subfigure}
    \caption{Statistics of Phase~\ref{phase2} methods for various types of NNs.}
    \label{fig:data12}
\end{figure}

According to types of the converted data, a specific NN model works better than the others. For example, CNN works well with images but does not work with raw text. From Figure ~\ref{fig:data1-3}, we observe that use of embedding for DNN (42.9\%), RNN (28.6\%) and LSTM (14.3\%) models approximates to 85\%. This observation indicates that embedding methods are commonly used to generate sequential input data for DNN, RNN and LSTM models. Also, we observe that one-hot encoded data are commonly used as input data for DNN (33.4\%), CNN (33.4\%) and LSTM (16.7\%) models. This observation indicates that one-hot encoding is one of common Phase~\ref{phase2} methods to generate numerical values for image and sequential input data because many raw input data for security problems commonly have the categorical features. We observe that the CNN (66.7\%) model uses the converted input data using the $Others$ methods to express the specific domain knowledge into the input data structure of NN networks. This is because general vector formats including graph, matrix and so on can also be used as an input value of the CNN model.

\begin{figure}[t!]
    \centering
    \begin{subfigure}[b]{0.95\textwidth}
        \centering
        
        \begin{tikzpicture}
            [
                pie chart,
                slice type={generation}{myblue}{crosshatch},
                slice type={detection}{myyellow}{vertical lines},
                slice type={classification}{mylightblue}{grid},
                pie values/.style={font={\small}},
                scale=1.3
            ]
    
                \pie{Embedding}{100/classification}
                \pie[xshift=2.2cm]%
                    {Parsing \& Embddding}{58.3/classification,41.7/detection}
                \pie[xshift=4.4cm]%
                    {One-hot encoding}{100/classification}
                \pie[xshift=6.6cm]%
                    {None}{60/classification,20/detection,20/generation}
                \pie[xshift=8.8cm]%
                    {Others}{66.7/classification,33.4/detection}

                \legend[shift={(0cm,-1cm)}]{{Data Generation}/generation}
                \legend[shift={(3cm,-1cm)}]{{Object Detection}/detection}
                \legend[shift={(6cm,-1cm)}]{{Classification}/classification}
            
        \end{tikzpicture}
        \caption{Output of NN over Phase~\ref{phase2} methods.}
        \label{fig:data1-5}
    \end{subfigure}
    \hfill
    \begin{subfigure}[b]{0.78\textwidth}
        \centering
        
        \begin{tikzpicture}
            [
                pie chart,
                slice type={embed}{mylightblue}{north west lines},
                slice type={parsing}{myyellow}{crosshatch},
                slice type={onehot}{myblue}{vertical lines},
                slice type={none}{mygreen}{grid},
                slice type={others}{myred}{dots},
                pie values/.style={font={\small}},
                scale=1.3
            ]
    
                \pie{Classification}{43.8/embed,21.9/parsing,18.8/onehot,9.4/none,6.3/others}
                \pie[xshift=2.2cm]%
                    {Object Detection}{71.5/parsing,14.3/none,14.3/others}
                \pie[xshift=4.4cm]%
                    {Data Generation}{100/none}

                \legend[shift={(-0.5cm,-1cm)}]{{Embedding}/embed, {Parsing \& Embddding}/parsing}
                \legend[shift={(2cm,-1cm)}]{{One-hot encoding}/onehot, {None}/none}
                \legend[shift={(4cm,-1cm)}]{{Others}/others}
            
        \end{tikzpicture}
        \caption{Phase~\ref{phase2} methods over output of NN.}
        \label{fig:data1-6}
    \end{subfigure}
    \caption{Statistics of Phase~\ref{phase2} methods for various output of NN.}
    \label{fig:data13}
\end{figure}
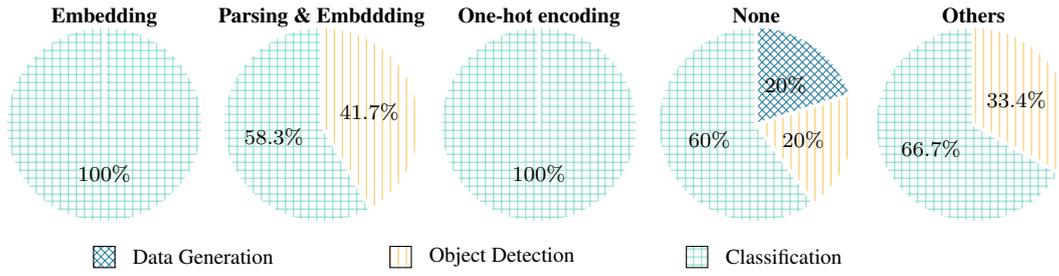
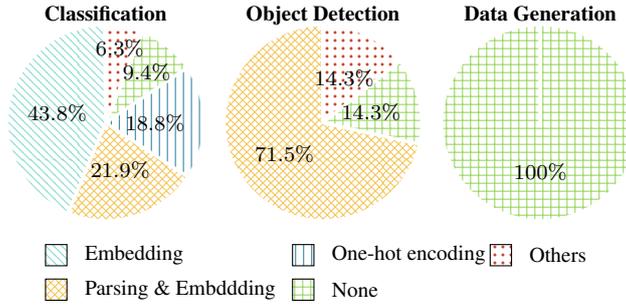

From Figure ~\ref{fig:data1-4}, we observe that DNN, RNN and LSTM models commonly use embedding, one-hot encoding and parsing combined with embedding. For example, we observe security papers of 54.6\%, 18.2\% and 18.2\% models use embedding, one-hot encoding and parsing combined with embedding, respectively. We also observe that the CNN model is used with various Phase~\ref{phase2} methods because any vector formats such as image can generally be used as an input data of the CNN model.

\subsubsection{Findings on output of NN models}

According to the relationship between output of security problem and output of NN, we may use a specific Phase~\ref{phase2} method. For example, if output of security problem is given into a class (e.g., normal or abnormal), output of NN should also be given into classification.

From Figure ~\ref{fig:data1-5}, we observe that embedding is commonly used to support a security problem for classification (100\%). Parsing combined with embedding is used to support a security problem for object detection (41.7\%) and classification (58.3\%). One-hot encoding is used only for classification (100\%). These observations indicate that classification of a given input data is the most common output which is obtained using Deep Learning under various Phase~\ref{phase2} methods. 

From Figure~\ref{fig:data1-6}, we observe that security problems, whose outputs are classification, commonly use embedding (43.8\%) and parsing combined with embedding (21.9\%) as the Phase~\ref{phase2} method. We also observe that security problems, whose outputs are object detection, commonly use parsing combined with embedding (71.5\%). However, security problems, whose outputs are data generation, commonly do not use the Phase~\ref{phase3} methods. These observations indicate that a specific Phase~\ref{phase2} method has been used according to the relationship between output of security problem and use of NN models.

\section{Further areas of investigation}
\label{sec:fur}
Since any Deep Learning models are stochastic, each time the same Deep Learning model is fit even on the same data, 
it might give different outcomes. 
This is because deep neural networks use random values such as random initial weights. 
However, if we have all possible data for every security problem, we may not make random predictions. 
Since we have the limited sample data in practice, we need to get the best-effort prediction results using the given Deep Learning model, 
which fits to the given security problem. 

How can we get the best-effort prediction results of Deep Learning models for different security problems? 
Let us begin to discuss about the stability of evaluation results for our selected papers for review. 
Next, we will show the influence of security domain knowledge on prediction results of Deep Learning models.
Finally, we will discuss some common issues in those fields.

\subsection{How stable are evaluation results?}
When evaluating neural network models, Deep Learning models commonly use three methods: train-test split; train-validation-test split; and $k$-fold cross validation. A train-test split method splits the data into two parts, i.e., training and test data. Even though a train-test split method makes the stable prediction with a large amount of data, predictions vary with a small amount of data. A train-validation-test split method splits the data into three parts, i.e., training, validation and test data. Validation data are used to estimate predictions over the unknown data. $k$-fold cross validation has $k$ different set of predictions from $k$ different evaluation data. Since $k$-fold cross validation takes the average expected performance of the NN model over $k$-fold validation data, the evaluation result is closer to the actual performance of the NN model.

From the analysis results of our selected papers for review, we observe that 40.0\% and 32.5\% of the selected papers are measured using a train-test split method and a train-validation-test split method, respectively. Only 17.5\% of the selected papers are measured using $k$-fold cross validation. This observation implies that even though the selected papers show almost more than 99\% of accuracy or 0.99 of F1 score, most solutions using Deep Learning might not show the same performance for the noisy data with randomness.

To get stable prediction results of Deep Learning models for different security problems, we might reduce the influence of the randomness of data on Deep Learning models. At least, it is recommended to consider the following methods:

\begin{itemize}
    \item \textbf{Do experiments using the same data many time}: To get a stable prediction with a small amount of sample data, we might control the randomness of data using the same data many times.
    \item \textbf{Use cross validation methods, e.g. $k$-fold cross validation}: The expected average and variance from $k$-fold cross validation estimates how stable the proposed model is.
\end{itemize}

\subsection{How does security domain knowledge influence the performance of security solutions using Deep Learning?}

When selecting a NN model that analyzes an application dataset, e.g., MNIST dataset~\cite{MNIST-2010}, we should understand that the problem is to classify a handwritten digit using a $28\times28$ black. Also, to solve the problem with the high classification accuracy, it is important to know which part of each handwritten digit mainly influences the outcome of the problem, i.e., a domain knowledge.

While solving a security problem, knowing and using security domain knowledge for each security problem 
is also important due to the following reasons (we label the observations and indications that realted to domain knowledge with `$*$'): 
{\color{myblack}
Firstly, \textit{the dataset generation, preprocess and feature selection highly depend on domain knowledge.} 
Different from the image classification and natural language processing, 
raw data in the security domain cannot be sent into the NN model directly. 
Researchers need to adopt strong domain knowledge to generate, extract, or clean the training set. 
Also, in some works, domain knowledge is adopted in data labeling
because labels for data samples are not straightforward.

Secondly, \textit{domain knowledge helps with the selection of DL models and its hierarchical structure.}
For example, the neural network architecture (hierarchical and bi-directional LSTM) designed in DEEPVSA~\cite{guo2019deepvsa} 
is based on the domain knowledge in the instruction analysis.

Thirdly, \textit{domain knowledge helps to speed up the training process.}
For instance, by adopting strong domain knowledge to clean the training set, 
domain knowledge helps to spend up the training process while keeping the same performance. 
However, due to the influence of the randomness of data on Deep Learning models, 
domain knowledge should be carefully adopted to avoid potential decreased accuracy. 

Finally, \textit{domain knowledge helps with the interpretability of models' prediction.}
Recently, researchers try to explore the interpretability of the deep learning model in security areas,
For instance, LEMNA~\cite{guo2018lemna} and EKLAVYA~\cite{chua2017neural}
explain how the prediction was made by models from different perspectives.
By enhancing the trained models‘ interpretability, they can improve their approaches' accuracy and security.
The explanation for the relation between input, hidden state, and the final output is based on domain knowledge.

\subsection{Common challenges}
In this section, we will discuss the common challenges when applying DL to solving security problems. 
These challenges as least shared by the majority of works, if not by all the works.
Generally, we observe 7 common challenges in our survey:
\begin{enumerate}
    \item The raw data collected from the software or system usually contains lots of noise. 
    \item The collected raw is untidy. For instance, the instruction trace, the 
    Untidy data: variable length sequences, 
    \item Hierarchical data syntactic/structure. As discussed in Section~\ref{sec:pa:dis}, 
    the information may not simply be encoded in a single layer, rather, it is encoded hierarchically, and the syntactic is complex.
    \item Dataset generation is challenging in some scenarios. Therefore, the generated training data might be less representative or unbalanced.  
    \item Different for the application of DL in image classification, and natural language process, which is visible or understandable, the relation between data sample and its label is not intuitive, and hard to explain.
\end{enumerate}

\subsection{Availability of trained model and quality of dataset.}
Finally, we investigate the availability of the trained model and the quality of the dataset.
Generally, the availability of the trained models affects its adoption in practice,
and the quality of the training set and the testing set will affect the credibility of testing results and comparison between different works. 
Therefore, we collect relevant information to answer the following four questions and shows the statistic in Table~\ref{tab:dataset}:
\begin{enumerate}
    \item Whether a paper's source code is publicly available?
    \item Whether raw data, which is used to generate the dataset, is publicly available?
    \item Whether its dataset is publicly available?
    \item How are the quality of the dataset?
\end{enumerate}

    \begin{table*}[htbp] 
            \centering
            \begin{threeparttable}
            \centering
            \caption{Analysis of the datasets and trained model.}
            \label{tab:dataset}
            \footnotesize
            \begin{tabular}{cc|p{1.5cm}<{\centering}p{1.5cm}<{\centering}p{1.5cm}<{\centering}cc}
            \hline\hline
            \textbf{Topic} & \textbf{Paper} & \textbf{Source Available} & \textbf{Raw Data Available}\tnote{1}& \textbf{Dataset Available}\tnote{2} & \multicolumn{2}{c}{\textbf{Quality of Dataset}}\\
                \cline{6-7}& & & & & Sample Num & Balance \\
               
                \hline
                \multirow{2}{*}{PA}& RFBNN~\cite{shin2015recognizing}  & \ding{55} &  \ding{51}   & \ding{55} & N/A & N/A \\ 
                \cline{3-7}
                                    &EKLAVYA~\cite{chua2017neural}  & \ding{55} &  \ding{51}   & \ding{55} & N/A & N/A \\ 
                \hline
                \multirow{2}{*}{ROP}& ROPNN~\cite{li2018ropnn}  & \ding{55} &  \ding{55}   & \ding{55} & N/A & N/A \\ 
                \cline{3-7}
                                    & HeNet~\cite{chen2018henet}  & \ding{55} &  \ding{55}   & \ding{55} & N/A & N/A \\ 
                \hline
                \multirow{2}{*}{CFI}& Barnum~\cite{yagemann2019barnum}  & \ding{51} &  \ding{55}   & \ding{55} & N/A & N/A \\ 
                \cline{3-7}
                                    & CFG-CNN~\cite{phan2017convolutional}  & \ding{51} &  \ding{51}   & \ding{55} & N/A & N/A \\
                \hline
                \multirow{2}{*}{Network}& 50b(yte)-CNN\cite{millar2018deep}  & \ding{55} &  \ding{51}   & \ding{55} & 115835 & \ding{55} \\ 
                \cline{3-7}
                                    & PCCN~\cite{zhang2019pccn}  & \ding{51} &  \ding{51}   & \ding{51} & 1168671 & \ding{55} \\    
                \hline
                \multirow{2}{*}{Malware}& Rosenber~\cite{Rosenberg2018}  & \ding{55} &  \ding{55}   & \ding{55} & 500000 & \ding{51} \\ 
                \cline{3-7}
                                    & DeLaRosa~\cite{DeLaRosa2018}  & \ding{55} &  \ding{55}   & \ding{55} & 100000 & \ding{55} \\        
                \hline
                \multirow{2}{*}{LogEvent}& DeepLog~\cite{Du:2017:deeLog}  & P\tnote{3} & \ding{51} &   P & N/A & \ding{55} \\ 
                \cline{3-7}
                                    & LogAnom~\cite{loganomaly:2019}  & \ding{55} &  \ding{51}   & P & N/A & \ding{55} \\ 
                \hline
                \multirow{2}{*}{MemoryFoensic}& DeepMem~\cite{Song:2018:DeepMem}  & \ding{51} &  \ding{51}   & \ding{51} & N/A & \ding{55} \\ 
                \cline{3-7}
                                    & MDMF\cite{Petrik:2018}  & \ding{55} &  \ding{55}   & \ding{55} & N/A & \ding{55} \\ 
                \hline
                \multirow{2}{*}{FUZZING}& NeuZZ~\cite{shi2019neuzz}  & \ding{51} &  \ding{55}   & \ding{55} & N/A & N/A \\ 
                \cline{3-7}
                                    & Learn \& Fuzz~\cite{godefroid2017learn}  & \ding{55} &  \ding{55}   & \ding{55} & N/A & N/A \\ 
                \hline
                \end{tabular}
                \begin{tablenotes}
                    \scriptsize
                    \item[1] ``Raw data'' refers to the data that used to generate training set but cannot be feed into the model directly. For instance, a collection of binary files is raw file. 
                    \item[2] ``Dataset'' is the collection of data sample that can be feed in to the DL model directly. For instance, a collection of image, sequence.
                    \item[3] ``P'' denotes that its source code or dataset is partially available to public.
                \end{tablenotes}
            \end{threeparttable}
        \end{table*}

We observe that both the percentage of open source of code and dataset in our surveyed fields is low,
which makes it a challenge to reproduce proposed schemes, make comparisons between different works, and adopt them in practice.
Specifically, the statistic shows that 1) the percentage of open source of code in our surveyed fields is low, only 6 out of 16 paper published their model’s source code.
2) the percentage of public data sets is low. Even though, the raw data in half of the works are publicly available, only 4 out of 16 fully or partially published their dataset. 
3) the quality of datasets is not guaranteed, for instance, most of the dataset is unbalanced. 
 
The performance of security solutions even using Deep Learning might vary according to datasets. 
Traditionally, when evaluating different NN models in image classification, standard datasets such as MNIST for recognizing handwritten 10 digits and 
CIFAR10~\cite{CIFAR-10-2010} for recognizing 10 object classes are used for performance comparison of different NN models. 
However, there are no known standard datasets for evaluating NN models on different security problems. 
Due to such a limitation, we observe that most security papers using Deep Learning do not compare the performance of different security solutions even when they consider the same security problem. 
Thus, it is recommended to generate and use a standard dataset for a specific security problem for comparison.
In conclusion, we think that there are three aspects that need to be improved in future research:
\begin{enumerate}
    \item Developing standard dataset. 
    \item Publishing their source code and dataset. 
    \item Improving the interpretability of their model.
\end{enumerate}
}

\section{Conclusion}
\label{sec:con}

This paper seeks to provide a dedicated review of the very recent research works on using Deep Learning techniques to solve computer security challenges. In particular, the review covers eight computer security problems being solved by applications of Deep Learning: security-oriented program analysis, defending ROP attacks, achieving CFI, defending network attacks, malware classification, system-event-based anomaly detection, memory forensics, and fuzzing for software security.  Our observations of the reviewed works indicate that the literature of using Deep Learning techniques to solve computer security challenges is still at an earlier stage of development.   

\section{Availability of data and materials}
Not applicable.
\section{Funding}
This work was supported by 
ARO W911NF-13-1-0421 (MURI), 
NSF CNS-1814679, and ARO W911NF-15-1-0576.
\section{Acknowledgements}
We are grateful to the anonymous reviewers for their useful comments and suggestions. 
\bibliographystyle{unsrt}
\bibliography{reference}

\end{document}